\documentclass[sigconf]{acmart}

\AtBeginDocument{%
  \providecommand\BibTeX{{%
    \normalfont B\kern-0.5em{\scshape i\kern-0.25em b}\kern-0.8em\TeX}}}

\copyrightyear{2023}
\acmYear{2023}
\setcopyright{acmlicensed}\acmConference[IMC '23]{Proceedings of the 2023
ACM Internet Measurement Conference}{October 24--26, 2023}{Montreal, QC,
Canada}
\acmBooktitle{Proceedings of the 2023 ACM Internet Measurement Conference
(IMC '23), October 24--26, 2023, Montreal, QC, Canada}
\acmPrice{15.00}
\acmDOI{10.1145/3618257.3624828}
\acmISBN{979-8-4007-0382-9/23/10}

\usepackage{balance}
\usepackage{pdflscape}
\usepackage{algpseudocode}
\usepackage{algorithm}
\usepackage{comment}
\usepackage{pdflscape}
\usepackage{xcolor}
\usepackage{float}
\usepackage{balance}
\usepackage{xspace}
\usepackage{subcaption}
\usepackage{multirow}
\usepackage{placeins}
\usepackage{placeins}
\usepackage{cleveref}
\usepackage{enumitem}
\usepackage{xurl}
\usepackage[normalem]{ulem}
\usepackage[framemethod=TikZ]{mdframed}
\usepackage[utf8]{inputenc}

\newcommand{\ndevices}{15\xspace}
\newcommand{\rtpml}{RTP ML\xspace}
\newcommand{\ipnonml}{IP/UDP Heuristic\xspace}
\newcommand{\ipml}{IP/UDP ML\xspace}
\newcommand{\rtpnonml}{RTP Heuristic\xspace}

\newcommand{\meet}{Meet\xspace}
\newcommand{\teams}{Teams\xspace}
\newcommand{\webex}{Webex\xspace}

\newcommand{\rev}[2]{#2}

\pgfsetarrows{latex-latex}
\tikzset{%
  base/.style = {inner sep=5pt,
                 text centered,
                 thin,
                 font=\rmfamily},
  round/.style = {base,
                  rectangle,
                  rounded corners=1ex,
                  draw=black,
                  fill=gray!20,
                  minimum height=0.35in}
}

\usepackage[all=normal,wordspacing]{savetrees}

\title{Estimating WebRTC Video QoE Metrics Without Using
Application Headers}

\author{Taveesh Sharma}
\email{taveesh@uchicago.edu}
\affiliation{%
  \institution{University of Chicago}
 \country{USA}
}
\author{Tarun Mangla}
\email{tmangla@iitd.ac.in}
\affiliation{%
  \institution{IIT Delhi}
  \country{India}
}
\author{Arpit Gupta}
\email{arpitgupta@ucsb.edu}
\affiliation{%
  \institution{UCSB}
 \country{USA}
}\author{Junchen Jiang}
\email{junchenj@uchicago.edu}
\affiliation{%
  \institution{University of Chicago}
 \country{USA}
}\author{Nick Feamster}
\email{feamster@uchicago.edu}
\affiliation{%
  \institution{University of Chicago}
 \country{USA}
}

\renewcommand{\shortauthors}{Taveesh Sharma, Tarun Mangla, Arpit Gupta, Junchen Jiang, \& Nick Feamster}
\renewcommand{\paragraph}[1]{\vspace*{0.03in}\noindent\textbf{#1}}

\begin{document}

\begin{sloppypar}
  \begin{abstract}
  The increased use of video conferencing applications (VCAs) has made
  it critical to understand and support end-user quality of experience
  (QoE) by all stakeholders in the VCA ecosystem, especially network
  operators, who typically do not have direct access to client software.
  Existing VCA QoE estimation methods use passive measurements of
  application-level Real-time Transport Protocol (RTP) headers. However, 
  a network operator does not
  always have access to RTP headers, particularly when VCAs use
  custom RTP protocols (e.g., Zoom) or due to system constraints (e.g.,
  legacy measurement systems). Given this challenge, this paper
  considers the use of more standard features in the network traffic, namely
  the IP and UDP headers, to provide per-second estimates of key VCA QoE
  metrics such as frame rate and video resolution. We develop a method
  that uses machine learning with a combination of flow statistics (e.g.,
  throughput) and features derived based on the mechanisms used by the
  VCAs to fragment video frames into packets. We evaluate our method for
  three prevalent VCAs running over WebRTC: Google Meet, Microsoft Teams,
  and Cisco Webex. Our evaluation consists of 54,696 seconds of VCA data
  collected from both (1) controlled in-lab network conditions and (2)
   \ndevices real-world access networks. We show that
  our approach yields similar accuracy compared to the
  RTP-based baselines despite using only IP/UDP data. For instance, we
can estimate frame rate within 2 FPS for up to 83.05\% of one-second intervals in
the real-world data, which is only 1.76\% lower than using the  RTP headers.

\end{abstract}



  \maketitle

  \section{Introduction} \label{sec:intro}



As users continue to depend on video conferencing applications (VCAs) for
remote participation in work, education, healthcare, and recreation, ensuring
a high quality of experience (QoE) when using VCAs is critical.  Although QoE
depends to some degree on the specific circumstances of end users, network
operators can often play an important role in mitigating QoE degradation
resulting from poor local network conditions.  A network operator who can
observe a VCA's QoE metrics may be able to diagnose and react to QoE
degradation, potentially preventing even transient congestion events from
affecting user experience.  Unfortunately, network operators lack direct
access to application QoE, and must infer QoE from the encrypted application
traffic as it traverses the network.  Methods exist to infer QoE from
video-on-demand applications, but these methods do not apply to inferring QoE
for VCAs, which is a different problem.  An important distinction
between VCAs and video-on-demand applications is that video-on-demand 
applications react to delay or loss by relying on a large playout buffer (i.e., of at least a few seconds); on the
other hand, VCAs must keep a short jitter buffer (specifically, less than
100~ms) and thus are susceptible to a wide range of incidents that can disrupt
or degrade network quality.

In this paper, we explore \textit{how to enable network operators to
  infer objective VCA QoE metrics at a per-second time granularity from passive
  measurements of network traffic}.  QoE is inherently
subjective~\cite{hossfeld2016formal}, making it challenging to infer on a
large scale, even for service providers, let alone network operators who have
no data from the instrumentation of the client, which can be useful for
directly inferring the user
experience. To address this challenge, objective application metrics are
commonly employed as a substitute for subjective QoE. The precise relationship
between these application-level metrics and user QoE can be determined through
user studies or data-driven methods~\cite{banitalebi2015effect} -- this is
complementary to the estimation of objective application metrics and is out of
scope of this paper. Furthermore, although VCA performance is determined by
both audio and video, past work has extensively examined audio QoE as a
function of network quality of service metrics~\cite{chen2006quantifying,
  Aggarwal2014}. Our primary focus, therefore, is to infer objective
metrics (described in Section~\ref{sec:background}) that
impact VCA video quality.

Recent work has proposed data-driven techniques, often leveraging machine
learning, to estimate VCA QoE metrics from network-layer
metrics~\cite{yan2017enabling,nikravesh2016qoe, carofiglio2021characterizing}.
However, most of these studies assume the ability to parse application-level
headers, which is not always the case.  Some VCAs, like Zoom, use proprietary
application protocols, posing challenges for extracting information using
standard network monitors~\cite{zoom_rtp}. In recent work, Michel et
al.~\cite{michel2022enabling} develop a method to detect Zoom application
traffic and extract encapsulated application headers. Yet, the proposed
approach will not work if Zoom changes its protocol format (e.g., if it
starts using a more complex encapsulation mechanism in the future). Moreover,
application headers are encrypted in certain scenarios, such as when
traffic is routed over a virtual private network (VPN), and it is likely that
all application headers will eventually be encrypted even for regular
traffic~\cite{uberti2023rfc}. Thus, \textit{this paper proposes methods to
  estimate video QoE using more standard features of the network traffic,
  specifically only IP/UDP headers}. A notable advantage of using IP/UDP headers
is that existing network monitoring systems can readily extract such
information at scale~\cite{star-flow}.

The QoE inference method we develop uses the semantics of video delivery in
VCA network protocols: Due to VCAs' real-time nature, each video frame is
encoded and transmitted immediately. These transmission characteristics give
rise to packet sizes and inter-arrival times containing important signal
about various QoE metrics, such as frame rate. By leveraging these insights,
we develop a heuristic as well as 
a machine learning-based model that estimates VCA QoE metrics at a fine time
granularity.  We evaluate our approach on three popular VCAs (Meet, Teams, and
Webex) that use WebRTC, an open-source framework providing real-time
communication capabilities to browsers and smartphones~\footnote{We focus on
  WebRTC-based VCAs as WebRTC provides mechanisms to collect ground truth QoE
  metrics, which are essential to evaluate the method we have developed. Our approach, however,
  applies to all VCAs that use Real-time Transport
  Protocol (RTP)}. To evaluate our approach, we collect data from in-lab
under diverse emulated network conditions as well as from \ndevices households
spanning different ISPs and speed tiers over a period of two weeks. Our
evaluation demonstrates that the proposed method achieves high accuracy in
estimating video QoE metrics for VCAs.

We make the following contributions:

\begin{itemize}[leftmargin=\parindent,align=left,labelwidth=\parindent,labelsep=0pt,topsep=0pt,itemsep=0pt]

  \item We develop a machine learning-based method that uses features
        informed by mechanisms used by VCAs to fragment a frame into
        packets and infer VCA QoE
        metrics at finer time \rev{granularities}{granularity} using only the IP/UDP headers.

  \item We develop an automated browser-based, VCA data collection framework
        and use it to evaluate our approach by collecting data under controlled
        in-lab network conditions as well as data from \ndevices households spanning a
        variety of ISP and speed tiers over two weeks. Both the code and
        data from the paper have been made public~\cite{githubrepo}.

  \item We demonstrate that using only IP/UDP headers can yield frame rate estimates
        within 1.50 frames of the ground truth QoE on average. To put it in perspective, we also
        compare accuracy using RTP headers, 1.33 of the ground
        truth QoE on average, a
        difference of only 0.17 frames.

  \item We show that a predictive model trained on data from controlled lab settings
  transfers to real-world networks. More specifically, the model transfers
  with a marginal drop in accuracy for two out of three VCAs.
  Furthermore, we characterize the network conditions under which
  the model has high errors and the potential reasons for
  errors.
\end{itemize}

  \section{Problem Context}\label{sec:background}

We provide background on video conferencing applications, the QoE
metrics, and detail the QoE inference problem.

\subsection{Video Conferencing Applications}

VCAs typically use Real-Time Transport Protocol
(RTP)~\cite{rfc3550} for sending audio and video data and Real-Time Transport
Control Protocol (RTCP)~\cite{huitema2003rfc3605} for control traffic. Although
VCAs can independently implement each of these protocols in the application,
the WebRTC open-source real-time communication framework has become extremely
prevalent, as it is supported by most modern browsers and devices (e.g.,
Android). We focus on WebRTC-based VCAs.

\paragraph{QoE metrics.} \rev{}{We focus on inferring objective metrics
pertaining to the video quality of conferencing. More specifically}, we focus
on the following four metrics: (1)~\textit{Video bitrate, }\rev{the total number of
video bits transmitted per second}{ defined as the total number of bits
received per second, with a lower bitrate indicating lower video quality.};
(2)~\textit{Frame rate}, \rev{}{defined as} the number of video
frames the application receives per second. \rev{}{A low frame rate leads to
reduced smoothness and realism of viewing experience} ; (3)~\textit{Frame
jitter} calculated as the standard deviation of the time gaps between
consecutive frames or inter-frame delay. \rev{}{A high frame jitter also
  affects the
smoothness of video playback, resulting in a jerky playback}. and
(4) \textit{Resolution}, the number of pixels in a video frame, with lower
resolution indicating lesser details in the video.

Additional metrics can affect a VCA's QoE, including end-to-end network
latency, as well as the resulting quality of the
audio~\cite{garcia2019understanding}. End-to-end network latency can be
challenging to measure from a single vantage point for UDP-based traffic;
previous work already estimates audio QoE for VoIP~\cite{chen2006quantifying}.


\subsection{Inference Problem} \label{sec:qoe_metrics}

\paragraph{Problem Statement.} We take as input a sequence of packets collected
from access nodes (e.g., border router) and
output the desired QoE metrics at a $W$-second granularity. The choice of $W$
ultimately depends on the network operator's ability to react to the inferred
QoE \rev{degradations}{degradation} by, for example, reconfiguring the network to mitigate the
inferred QoE degradation incidents. We also assume that the input consists only
of RTP packets from the VCA and contains no other traffic. We can safely make
this assumption because previous work has developed traffic classification
methods to identify packets associated with a specific VCA session~\cite{nistico2020comparative}.

\paragraph{Measurement Context.} We consider the case when operators use only
IP and UDP headers. This scenario is motivated by several observations: First,
for some VCAs that use non-standard versions of RTP (e.g., the native Zoom
client~\cite{zoom_rtp}), network operators do not have access to RTP headers
as these VCAs. Second, as has transpired with many other applications and
protocols (e.g., DNS~\cite{borgolte2019dns}, TLS~\cite{chai2019importance}),
we expect VCAs to encrypt the RTP headers in the future. Finally, extracting
IP and UDP headers is much more efficient and scalable than extracting RTP
headers; in fact, many existing network monitoring systems~\cite{star-flow}
already support extracting IP/UDP headers along with packet sizes and times.

  \section{Method} \label{sec:method}
In this section, we describe our QoE estimation method that uses only IP/UDP
headers. We assume access to traffic from a single VCA session and
\rev{comprise of}{it consists of} two steps.  The first
step involves isolating the video traffic from the audio component. Given the
distinct transmission techniques (e.g., encoding, error control) used for
audio and video, it becomes important to differentiate audio and
video packets. Once the video traffic is identified, the second step involves
using information from this traffic to infer the video QoE metrics.
We first describe these two steps for our method. This is followed by a
description of RTP baselines used for comparison.

\begin{figure}[t!]
	\centering
	\includegraphics[width=0.85\linewidth]{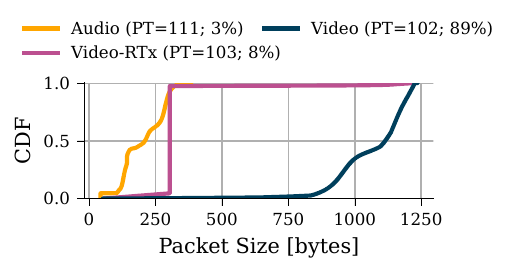}
	\caption{Packet sizes vs payload type for \teams.}
	\label{fig:teams_packetSizeCDF}
\end{figure}

\subsection{Media Classification}
\label{sec:media_classification}
Past work to distinguish media type relies on RTP headers~\cite{zoom_rtp,
	nistico2020comparative}. More specifically, a seven-bit RTP header
called \textit{payload type} can be used to identify the payload format.
For example, in case of \teams, we observe three different payload
types~(PT): (1).  PT = 111 for audio encoded using OPUS,  (2) PT=102 for
video encoded using H.264, and (3) PT = 103 for video retransmissions.
However, with no access to RTP headers, it becomes challenging to
identify the media type of an RTP packet. 

To overcome this challenge, we use the insight that voice samples can be
encoded in fewer bits than images. As a result, the audio packets are
typically smaller than video packets. Figure~\ref{fig:teams_packetSizeCDF} 
illustrates this phenomenon, showing the CDF of packet sizes
corresponding to audio, video, and video retransmissions from $16528$
seconds of Teams calls (see Section~\ref{sec:dataset} for details). The
actual packet media type is identified using the \textit{RTP Payload
	Type} header. The audio packet sizes range between $[89, 385]$ bytes; the
video packets are significantly larger, with 99\% of packets being
larger than $564$ bytes. Among video retransmissions, which constitute
 8\% of video packets, we find a significant proportion (92\%) of
packets with a packet length of $304$. These are likely keep-alive
messages for the retransmission transport stream as retransmissions are typically only sent in the case of packet losses.
Because these packets do not contain any video payload, it makes sense
to filter them out from the QoE inference step. The remaining video
retransmission packets are significantly larger.

This characteristic allows us to use a size threshold denoted as $V_{min}$ to
identify video packets. Any packet with size greater than or equal to
$V_{min}$ is tagged as a \textit{video} packet, while the remaining packets
are not considered. The value of $V_{min}$ can be determined by inspecting a
\rev{small number of}{few} VCA traces collected in the lab. 

\subsection{QoE inference}

We develop two approaches to infer QoE metrics from video traffic using only
IP/UDP headers. The first approach, referred to as \ipnonml, utilizes
VCA video delivery semantics. We find that relying solely on the heuristic
approach can lead to errors, particularly under high network jitter and loss.
We thus propose a machine learning(ML)-based approach called \ipml that relies on
a combination of network features, including both statistics on network
traffic and features derived using insights from the \ipnonml. 

\subsubsection{Heuristic}
\label{sec:heuristic}

Because VCAs are real-time and low latency application, each video frame generated at the sender is transmitted
over the network as soon as it has been encoded. From the network perspective,
each frame comprises one or more RTP packets. The VCA client transmits these
packets immediately, without waiting for additional frames. As a result,
\textit{a VCA session can be abstracted as a sequence of video frames, with
each frame transmitted sequentially over a group of RTP packets separate from
other frames}. Identifying the video frame boundaries (by identifying frame
end time) and frame size can enable inference of key QoE metrics described in
Section~\ref{sec:background}. Past work has relied on using RTP headers to
identify frame boundaries~\cite{zoom_rtp}. Without access to the RTP headers,
it is challenging to identify the frame boundaries. 

\paragraph{Key Insights:} To identify frame boundaries using IP/UDP headers,
we use insights from the mechanisms that VCAs use to divide frames into
packets.
We first consider whether there are patterns in
packet inter-arrival times (IAT). A frame is packetized and
transmitted immediately, which leads to microbursts on the network,
causing the inter-departure times to be shorter for packets within
the frame as compared to packets across frames. Unfortunately, this insight
is challenging to apply reliably to determine frame boundaries as packet
timings can change when packets traverse along the network. Thus, the
patterns in the inter-departure times may not appear in the 
inter-arrival time (IAT) at the receiver.

\begin{figure}[t!]
	\centering
	\includegraphics[width=0.85\linewidth]{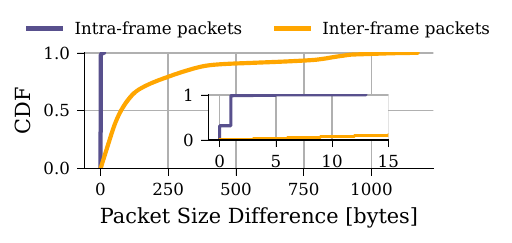}
	\caption{Intra- and inter-frame packet size difference for \teams}
	\label{fig:teamsPacketsizeDiff}
\end{figure}

We next consider whether there are unique patterns in packet sizes. An
advantage of using packet size is that it does not change during packet
transmission over the network. Interestingly, we find a unique pattern
in the packet sizes, i.e.,  packet sizes tend to resemble those within
the same frame and differ from packet sizes in consecutive frames. This
phenomenon occurs because VCAs typically fragment a frame into equal-sized packets. This is done
because the Forward Error Correction (FEC) mechanisms used to
protect against network losses are most bandwidth-efficient when
packets in a frame have equal length~\cite{rfc5109, rfc6184}.
Furthermore, due to dynamic nature of the underlying video content along
with variable bitrate encoding used by VCAs, consecutive frames exhibit
different sizes and, consequently different packet sizes.

Figure~\ref{fig:teamsPacketsizeDiff} illustrates this characteristic, showing the
CDF of size difference in consecutive intra-frame and inter-frame
packets, for more than 360,000 frames. The true frame boundaries are
identified based on the RTP timestamp header as explained in
Section~\ref{sec:rtp-baseline}.  For frames with more than two
packets, we show only the maximum size difference across all packets. The inter-frame size difference is the absolute size
difference between the first and the last packets of two consecutive
frames. We find that the intra-frame packet size difference is
less than two bytes for all but one packet. The inter-frame packet size
difference on the other hand is at least 2 bytes for more than 99.4\% of
the frames.

\paragraph{Frame boundary estimation}: Thus, we use a packet size difference threshold $\Delta_{size}^{max}$ and
declare frame boundary if the size difference between consecutive
packets is greater than $\Delta_{size}^{max}$. However, it is not sufficient
to compare only consecutive packets as packets can arrive out of order.
Therefore, instead of comparing with only the last packet, we
iteratively compare with up to $N^{max}$ packets that arrived before this
packet, beginning with the most recent packet. If the size difference of
the current packet is within $\Delta_{size}^{max}$ for any of these
packets, it is considered as part of the same frame as the matching
packet. Otherwise, the packet is assigned as a part of new frame. The
exact heuristic is described in Algorithm~\ref{algo:ip-heuristic} in
the Appendix. 

The parameters of the heuristic, i.e., $N^{max}$ and $\Delta_{size}^{max}
$, can be determined by inspecting few traces for a given VCA in the
lab. Intuitively, a large value of $N^{max}$ can account for all
out-of-order packet arrivals. However, it also increases the probability
of incorrectly combining a packet from a new frame to an earlier frame
with a similar size. Thus, the value of $N^{max}$ should be set
carefully. We analyze the sensitivity of the heuristic to
different values of $N^{max}$ in our evaluation.

\paragraph{QoE estimation from frames}: Once the frame boundaries have been identified, 
for a single session S, we \rev{obtaine}{obtain} a sequence of frames along with their sizes. We use this
information to estimate the key QoE metrics over a window $W$ of duration
$w$ seconds in the following manner: 

\begin{itemize}[leftmargin=\parindent,align=left,labelwidth=\parindent,labelsep=0pt,topsep=0pt,itemsep=0pt]
	\item \textbf{Video bitrate}: It is simply the time average of the total bits across all 
	frames transmitted in the window \textit{W}. 
	\item \textbf{Frame rate}: It is simply the number of frames transferred per second in 
	the window \textit{W}. More specifically, Frame rate = $\frac{\sum_{i=1}^{N} I(ET_{i} \in 
		\textit{W})}{w}$. Here, indicator function $I$ equals one if the frame end time is 
	within the window, and zero otherwise. 
	\item \textbf{Frame jitter}: It is calculated as the standard deviation of difference 
	in end times ($ET_{i}$ - $ET_{i-1}$) of consecutive frames received over the window 
	\textit{W}. 
\end{itemize}

  We do not estimate frame resolution
  using this method as there is no direct signal in the frame reflecting
  its resolution. Intuitively, one can design a machine learning-based method
  that uses frame sizes and FPS from the heuristic to predict video
  resolution. This, however, is similar in principle to the machine
  learning-based method described in Section~\ref{sec:machinelearning};
  hence, we skip implementing the approach for simplicity.

\begin{table*}[t!]
  \centering
  \small
\begin{tabular}{|l|l|}
\hline
\multicolumn{1}{|c|}{\textbf{Category}} &
\multicolumn{1}{c|}{\textbf{Features}}                                                    \\ \hline
\hline
Flow-level statistics                  & Bytes per second, packets per second,  packet size (5) and inter-arrival statistics (5)   \\ \hline
IP/UDP features based on VCA semantics               & \# unique packet sizes, \# microbursts                                                    \\ \hline
RTP Headers                           & \# unique RTP timestamps (4),
marker bit sum (1),  out-of-order sequence numbers (1), RTP lag (5) \\ \hline
\end{tabular}
\caption{Summary of features extracted from traffic. Numbers in
\rev{parenthesis}{parantheses} reflect the count of features. The \ipml approach uses the first two categories of features, while the \rtpml approach uses the first and third category of features. }
\label{tab:feat-summary}
\end{table*}

\subsubsection{Machine Learning Approach}
\label{sec:machinelearning}
\paragraph{Why use machine learning?}: The heuristic described in
Section~\ref{sec:heuristic} relies on assumptions that can break
under certain \rev{conidtions}{conditions}. For instance, under high latency jitter or
packet loss, packets can arrive out of order leading to incorrect
estimation of frame boundaries. Although we add parameters (e.g.,
use a packet lookback $N^{max}$ $>$ 1) that alleviate the errors to some
extent, it still does not completely solve the problem. More importantly,
there are other, complimentary, signals in the network data that can
inform QoE estimation. For instance, given the real-time nature of the
VCAs, throughput is a potential indicator of few QoE metrics such as
video bitrate. Including multiple such signals into a heuristic can
quickly make it complicated. Therefore, we consider a data-driven
approach that considers multiple features derived from the network data
along with supervised machine learning models. We now describe our
approach. 

\paragraph{Input features}: We use a common set of features to predict
all QoE metrics. The features considered can be divided into two
categories:

\begin{itemize}
[leftmargin=\parindent,align=left,labelwidth=\parindent,labelsep=0pt,topsep=0pt,itemsep=0pt]
  \item \paragraph{VCA semantics-based}: These include two features that are
    informed by how VCAs fragment frame into packets as
    described in Section~\ref{sec:heuristic}. The first feature is the number of unique packet sizes observed in the
    prediction window $W$. The second feature is the number of
    microbursts of packets in the prediction window $W$. A microburst is
    defined as a sequence of packets with the consecutive inter-arrival
    times within a threshold $\theta_{IAT}$. Therefore, the microburst
    count is simply the number of consecutive packets with inter-arrival
    time $\geq$ $\theta_{IAT}$. Intuitively, these features  can help inform the frame
    boundaries and consequently the key video QoE metrics. 
  \item \paragraph{Flow-level statistics}: We also derive a set of key
    statistics from the IP/UDP headers of video packets. These include
    number of bytes and packets per second as well as five
    statistics on packet sizes and inter-arrival times namely mean, standard
    deviation, median, minimum and maximum. \rev{Intutitively}{Intuitively}, given the
    real-time nature of VCAs,
    any transient \rev{degradations}{degradation} in the VCA QoE metrics would also
    be evident in one or more of these statistics. 
  \end{itemize}

  In total, we compute 14 features for each prediction window $W$ as
  summarized in Table~\ref{tab:feat-summary}.

\subsection{RTP Baselines}
\label{sec:rtp-baseline}
To benchmark the accuracy of our approach using IP/UDP headers, we
also consider two RTP-based approaches as baselines. The first approach is a
heuristic approach, called \rtpnonml, and the other is a machine
learning-based approach called \rtpml. We now describe both of these 
approaches.

\paragraph{\rtpnonml}: This is similar to the approach used by Michel et
al. to estimate QoE metrics for Zoom~\cite{michel2022enabling} and is based on the same
insight as the \ipnonml approach, i.e., a VCA session can be modeled as a sequence of frames. 
To identify frame boundaries, it uses the \textit{RTP timestamp} field from the packet headers. The \textit{RTP timestamp} is used to determine the correct order for
media playback, as well as to synchronize audio and video streams. Packets belonging to the same frame receive the same \textit{RTP
Timestamp}, and
thus the field can be used to identify frame boundaries. To detect the
end of frames, the approach also uses the \textit{Marker bit} in the RTP
header. This bit is set only for the last packet of each frame and is
used to detect the end of frames.

Using this approach, we can identify the
sequence of frames in the \rev{prediciton}{prediction} window $W$, along with frame
completion time and frame size. We then use similar method as
described in Section~\ref{sec:heuristic} to estimate frame rate, frame
jitter and bitrate. 

\paragraph{\rtpml}: This is similar to the \ipml approach and uses
machine learning-based \rev{methds}{methods} to estimate QoE metrics. The input
features, however, are derived from RTP headers. We
consider the following set of RTP-based features:

\begin{itemize}[leftmargin=\parindent,align=left,labelwidth=\parindent,labelsep=0pt,topsep=0pt,itemsep=0pt]
	\item \textbf{RTP timestamps:} We calculate the
	number of unique RTP timestamps over each stream individually as
	well as their intersection and union. 
\item \textbf{Marker bit sum:}  It is the sum of
	marker bit for all packets in the prediction window. We calculate
	this feature separately for video and retransmission streams. 
	\item \textbf{Number of out-of-order video sequence numbers:}  We calculate the
  total number of
	discontinuities in video packet RTP sequence numbers over the
  prediction window. It is
	used as a signal for packet re-ordering and loss.
  \item \textbf{RTP Lag:} It captures the delays in frame transmission.
    We assume that the first frame had zero delay. For each frame $i$, we calculate the transmission delay as the difference between 
    its \rev{receive}{reception} time $t_i$ and transmission time, which is calculated
    as $t_{0} + \frac{RTP_i - RTP_0}{SF}$. Here, $SF$ is the sampling
    frequency for generating RTP timestamps and is typically 90,000 for
    most video codecs~\cite{rfc6184}. We then calculate the five
    statistics across frame transmission delays.  
    \end{itemize}

In addition, we also use the flow-level
statistics as summarized in Table~\ref{tab:feat-summary}. This is done for similar reasons as
described for the \ipml approach.

  \section{Experiment Setup and Datasets}
\label{sec:dataset}
This section \rev{describe}{describes} our \rev{exprementation}{experimentation} framework and the different
datasets we use to evaluate our methodology.

We consider WebRTC-based
VCAs for evaluation as WebRTC is a popular framework used by most VCAs
for their browser version. \rev{Moreoever}{Moreover}, it is possible to obtain ground
truth QoE metrics for WebRTC-based VCAs using the
\texttt{webrtc-internals} API provided by Google
Chrome~\cite{webrtc-internals}. To collect data
for evaluation, we build an automated
browser-based framework that initiates calls for a given VCA over a
browser.  The framework uses PyAutoGUI, a UI automation framework, for starting and
ending the calls. We collect  data for
three popular VCAs, namely \meet, \teams, and \webex. The framework,
however, is extensible to other VCAs.

We conduct 2-person calls, each lasting for a variable
duration. For consistency, we use a virtual web camera at one of the
endpoints streaming a predefined short video on loop and log the
QoE metrics on the other endpoint. At the end of the call, we collect
both network traces and WebRTC logs.

\subsection{Matching ground truth with estimates.} We compare our QoE estimates
with per-second metrics reported by \texttt{webrtc-internals}. We match the two
datasets using the timestamp fields in the two datasets. The \textit{webrtc-internals} reports only the
start and end times of data collection. We assume that the reported per-second
metrics are collected at one-second interval; this matching approach may not be
perfect in certain cases, such as when WebRTC logs contain time intervals that
are slightly out of phase. To address this as much as possible, during
our analysis, we filter out logs
where we observe fewer per-second logs compared to the duration of the
call.

\subsection{Network Conditions}
\label{sec:network-conditions}
To evaluate under diverse network
conditions, we collect two kinds of data: (1). in-lab data under emulated network conditions, and (2).  data from \ndevices
households under real-world network conditions.

\paragraph{In-lab Data}
\label{sec:lab-data}
The data is collected by conducting calls between two machines in the
lab under emulated network conditions. We emulate dynamic network
conditions using the \texttt{tcp-info} stats dataset
from the Measurement Lab's Network Diagnostic Test (NDT), a public
dataset containing speed
tests taken by real users across the world~\cite{mlab-data}. The test measures TCP throughput by
flooding the link for ten seconds. We use the samples of instantaneous
throughput and RTT, called \texttt{tcp-info} stats, collected multiple
times during the test~\cite{tcp-info-mlab}. More specifically, we
emulate the same
sequence of RTT and packet loss values as observed in a single test, while the
throughput values are sampled from a normal distribution with the same
mean and variance as the test throughput. We did
not use the throughput samples directly as they include
throughput observed during the TCP slow-start period. Each throughput,
delay, and loss value is emulated for a period of 1 second. We only use
traces with average speeds below 10~Mbps to create
challenging network conditions. We collect around
$11$k~seconds, $15$k~seconds, and $13$k~seconds of \meet, \teams, and
\webex data, respectively. \rev{Figure~\ref{fig:qoe-cdf} shows the distribution of the three QoE metrics for
all three VCAs.}{As expected, we find differences in ground truth QoE metrics across the VCAs despite the presence of similar network conditions. For instance, the median bitrate is 500~kbps for \webex, whereas it is 1700~kbps for \teams (see Figure~\ref{fig:qoe-cdf} in Appendix for other metrics). These differences can be attributed to design variations within the VCAs. Conducting evaluation across multiple VCAs can help us understand the generalizability of our methodology.}

\paragraph{Real-world Data}. We note that the in-lab data is not a perfect emulation of the
real-world networks; therefore, we complement our data with real-world
VCA data. For this purpose, we deploy Raspberry Pi (RPi) devices in \ndevices
households, directly connected to the home router. These
households are recruited with the help from community organizations and
are located in a major city, spanning different neighborhoods, ISPs, and
speed tiers~\cite{sharma2022benchmarks, macmillan2023comparative}. Although our sample size is limited, it serves as an
additional
independent data source, capturing real-world network conditions,
which allows us to thoroughly test our methods.

The RPi collects VCA data by initiating a
15-25s call every 30 minutes to an end point located inside a
cloud network. The VCA is selected randomly from the three VCAs. \rev{}{The cloud endpoint and the RPi both join the VCA call as two different participants. }During the call, the video on the RPi is kept off while the cloud-network end point streams a
predefined video \rev{}{ over a virtual camera interface, same as in the lab experiments}. \rev{We turned the video off}{We do not stream video} on the
RPi as it increases the CPU utilization, leading to degradation in call quality due to non-network
reasons. For each call, we
log the ground truth QoE metrics and the network traffic on the
RPi and export the data to a centralized server at the end of the call.

The data collection spanned over a period of two weeks and includes 320 \meet calls, 178 \teams calls,
and 417 \webex calls. \rev{Figure~\ref{fig:qoe-cdf-rw} show the CDF of QoE metrics as obtained
from the WebRTC logs.}{} Compared to the in-lab data, the average QoE
metrics \rev{exhibhit}{exhibit} higher values \rev{, likely due to better
network conditions along the public cloud to RPi downstream path
compared to the in-lab emulated network conditions}{} \rev{}{(see Figure~\ref{fig:qoe-cdf-rw} in Appendix for the distribution)}. This improvement is expected as the download
speeds of access networks, likely to be the bottleneck in this case,
have significantly improved. We also, however, observe a small fraction
of calls with low QoE\rev{}{, indicating the presence of variability in the real-world network conditions}.

\subsection{Parameter Setting and Model Training}
The \ipnonml uses two parameters, $\Delta_{size}^{max}$ and $N^{max}$,
that are VCA-specific. We set these parameters by sampling \rev{a small
number of}{a few} sessions for each VCA. We use a value of 2 bytes for
$\Delta_{size}^{max}$ across all VCAs. The value of  $N^{max}$ is set
to 3, 2, and 1 for \meet, \teams, and \webex, respectively. For the ML methods, we use random forests as it was the most accurate
among the classical supervised machine learning models. The accuracy
numbers for these methods are reported over a 5-fold cross validation. 

For the ML methodology, \rev{we use random forests as it was the most accurate
among the classical supervised machine learning models.}{we experiment with several classical supervised ML models, specifically Support Vector Machines (SVMs), decision trees, and random forests. However, in this paper, we present the results obtained using only random forests, as they consistently yield the highest accuracy. This finding  aligns with prior research within the field that has leveraged ML-based techniques for network data analysis~\cite{farnaaz2016random, dimopoulos2016measuring, mangla2020drop, carofiglio2021characterizing}.} In addition, the accuracy numbers for ML-based techniques are reported after 5-fold cross-validation.
  \section{Evaluation} \label{sec:evaluation}

\begin{table}[h]
  \centering
  \small
  \begin{tabular}{|c|cc|c|}
    \hline
    \multirow{2}{*}{\textbf{Actual}} & \multicolumn{2}{c|}{\textbf{Predicted}} & \multirow{2}{*}{\textbf{Total}}           \\ \cline{2-3}
                                     & \multicolumn{1}{c|}{Non-Video}          & Video                           &         \\ \hline
    Non-video                        & \multicolumn{1}{c|}{98.3\%}             & 1.7\%                           & 67,830  \\ \hline
    Video                            & \multicolumn{1}{c|}{0\%}                & 100\%                           & 360,481 \\ \hline
  \end{tabular}\caption{Media classification accuracy for \meet}
  \label{tab:media_meet}
\end{table}

Our evaluation \rev{anaylzes}{analyzes} the  accuracy of IP/UDP methods, particularly in
comparison to the RTP baselines, using both in-lab and real-world
datasets. Furthermore, we examine the potential sources of errors as well as
identify the most important features of ML methods. Later, we analyze
the transferability of ML models, characterize the network conditions
where the models err, and quantify the impact of the prediction window on
model accuracy. 


\subsection{In-lab Data Results}
We describe the accuracy of our methods in classifying media and
estimating each QoE metric for in-lab data.

\subsubsection{Media Classification Accuracy}
\label{sec:media-classification}
Identifying video packets is a common step for both the IP/UDP
methods. The ground truth
is obtained by inspecting the \texttt{Payload Type} RTP Header.  Table~\ref{tab:media_meet} shows the normalized confusion
matrix for video packet identification for \meet. \rev{}{The}
accuracy of identifying video packets is generally high. However, a
small fraction of non-video packets get \rev{mis-classified}{misclassified}
as video. Upon closer inspection, we find that these \rev{misclassifed}{misclassified}
packets are server hello messages over DTLSv1.2 and the key exchanges at
the beginning of the call.


\paragraph{Impact of misclassification on QoE estimation}. For \ipnonml,
these \rev{additonal}{additional} packets can result in false frame boundaries, leading
to an overestimation of the number of frames. On the other hand, the \ipml
method may be more resilient to minor errors in video traffic
classification as it relies on multiple signals in the network traffic.

%
\begin{figure}[h]
  \begin{center}
    \includegraphics[width=0.75\linewidth]
    {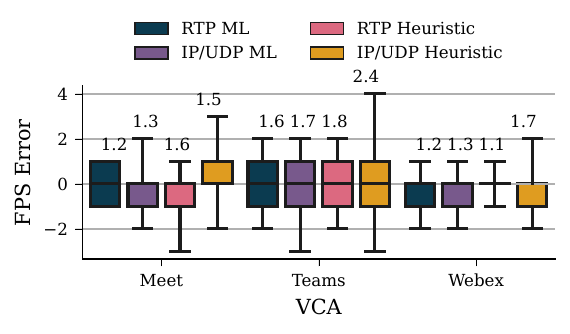}
  \end{center}
  \caption{Frame rate errors. The
    whiskers represent the $10^{th}$ and $90^{th}$ percentile values. The numbers
    represent the MAE.}
  \label{fig:cdf-fps}
\end{figure}

\subsubsection{Frame Rate} \label{sec:fps-lab}

Figure~\ref{fig:cdf-fps} shows the error distribution in frame rate along with the
Mean Absolute Error (MAE) values across VCAs. We observe a consistent trend	
in MAE values across all VCAs: \rtpml < \ipml < \rtpnonml < \ipnonml.
However, we observe a deviation from this trend in \webex, where the MAE
of the \rtpnonml
is lower than that of both \rtpml and \ipml approaches, and in \meet
where the MAE of the
\ipnonml is lower than that of \rtpnonml. Moreover, the MAE remains
within a 2 FPS margin in all
cases, except for the \ipnonml over \teams. 

In general, both heuristics tend to have higher errors than ML-based
methods. One potential reason for this could be that the WebRTC frame
rate is reported after accounting for additional application-level
delays, such as jitter buffer delays, which are not observable directly
from the network traffic. The ML-based methods trained on
application-level ground truth can potentially calibrate their
prediction to account for such mismatch, while that is not
possible for the two heuristics.

Interestingly, the \ipml method errors have a similar distribution
as \rtpml. \textit{This indicates that IP/UDP headers can estimate frame rate with
comparable accuracy to RTP headers}. In contrast, the \ipnonml has
the highest errors. This is surprising as we expect the \ipnonml to have similar 
accuracy as the \rtpnonml. We now examine the error causes for the
\ipnonml approach.

\paragraph{Why does the \ipnonml exhibit higher errors?}
The \ipnonml relies on the observation that the inter-frame packet size
difference is larger than the intra-frame packet size difference. However,
this is not true for a few cases:

\paragraph{Case 1.} If two consecutive frames are similar in
    size, it will end up combining them. \rev{We illustrate
    such a case in Figure~\ref{fig:teams-asgn}.}{}   

    \paragraph{Case 2.} If the packets within a frame have a size
    difference greater than $\Delta_{size}^{max}$, they will be split
    into multiple frames. We observe this primarily for \meet, where 
    a fraction of frames \rev{have}{contains} packets with large intra-frame packet size
    differences.  

    \paragraph{Case 3.} If packets arrive out-of-order, the frames will
    get interleaved. As a result, the \rev{heursitic}{heuristic} will
    create false frame \rev{bounderies}{boundaries} and overestimate the frame rate.

\begin{figure}[h]
  \begin{center}
    \includegraphics[width=0.85\linewidth]{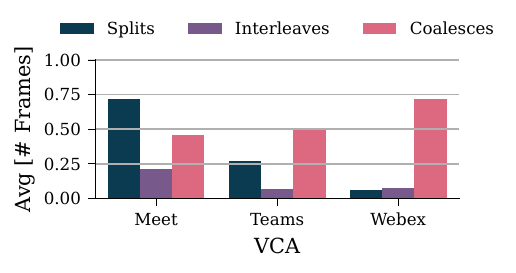}
  \end{center}
  \caption{Different types of errors in the inter- and intra-frame packet size difference assumption}
  \label{fig:meet-inacc}
\end{figure}

We analyze the frequency of each type of error in our data as shown in
Figure~\ref{fig:meet-inacc}. For \meet, we observe a greater number of
splits for about 0.72 frames in one prediction window on average,
leading to overestimation (see Figure~\ref{fig:cdf-fps}). We detect
these splits by calculating the number of frames where the intra-frame
packet size is greater than $\Delta_{size}^{max}$.
In Figure~\ref{fig:meet-inacc}, we also see that a higher
percentage of erroneous coalesces leads to underestimation of FPS in
Webex. We calculate these by estimating the number of frames to which
more than one RTP timestamp was assigned by the \ipnonml in the prediction window.

\begin{figure}[h]
  \begin{center}
    \includegraphics[width=0.75\linewidth]{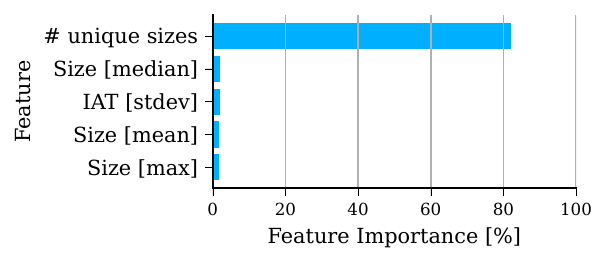}
  \end{center}
  \caption{Top 5 features' importance scores for \ipml frame rate predictions for Teams}
  \label{fig:fps-teams-lab}
\end{figure}

\paragraph{Feature importance for \ipml method}. Figure~\ref{fig:fps-teams-lab} 
shows the top 5 features
for frame rate prediction in the case of \teams. We observe a high feature
importance for the \textit{\# unique sizes} feature. We also observe the
significant importance of this feature for \meet and \webex (see
Figure~\ref{fig:fps-fimp} \rev{}{in the Appendix}). 
The \rev{prevalance}{prevalence} of \textit{\# unique sizes} among the top 5 features of
all VCAs suggests a strong correlation between frame rate and unique
packet sizes, enabling accurate frame prediction even without utilizing
the RTP headers.

Notably, the other semantic-based feature, \textit{\#
  microbursts}, does not appear among the top 5 features. This suggests
that there is a significant distortion of inter-packet times along the
network path. Furthermore, an ML approach, like \ipml, can take advantage of
other signals in the network, which is absent in the \ipnonml. For
example, the most important feature is \textit{IAT [min]} for \meet and
\textit{\# bytes} for \webex.

\begin{figure}[t!]
  \centering
  \begin{subfigure}{0.85\linewidth}
    \centering
    \includegraphics[width=\linewidth]
    {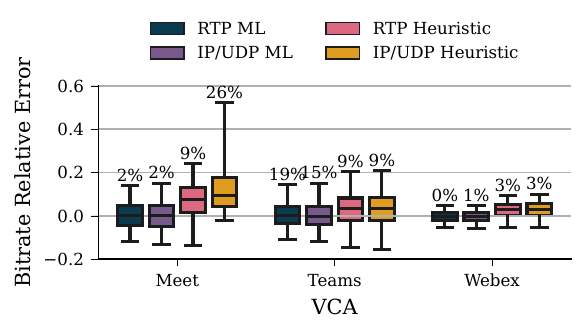}
    \caption{Bitrate}
    \label{fig:bitrate_boxplot_lab}
  \end{subfigure}
  \begin{subfigure}{0.85\linewidth}
    \centering
    \includegraphics[width=\linewidth]{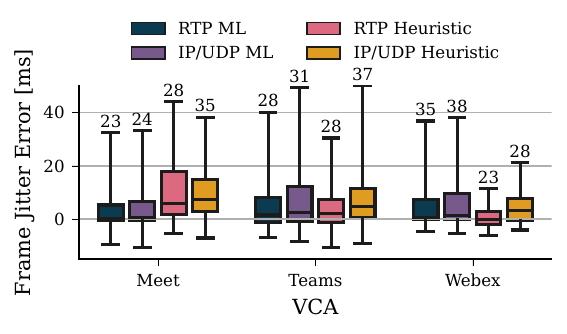}
    \caption{Frame jitter}
    \label{fig:jitter_error}
  \end{subfigure}
  \caption{Distribution of errors across the VCAs. The
    whiskers represent $10^{th}$ and $90^{th}$ percentile values. The numbers
    represent the MRAE for bitrate and MAE for frame jitter.}
  \label{fig:bitrate-jitter-error}
\end{figure}

\begin{figure}[h]
  \begin{center}
    \includegraphics[width=0.75\linewidth]{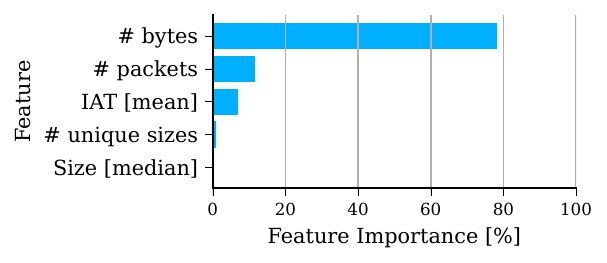}
  \end{center}
  \caption{Top-5 features along with feature importance scores for bitrate estimation using the IP/UDP ML method for Webex}
  \label{fig:fimp-bitrate-webex-ip}
\end{figure}

\subsubsection{Bitrate} \label{sec:bitrate-lab}

We calculate the relative bitrate error, defined as the ratio of bitrate
error and the ground truth bitrate. Using relative
values facilitates the comparison of errors across VCAs, especially because
the ground truth bitrate distributions differ significantly across VCAs.  Figure~\ref{fig:bitrate_boxplot_lab} shows the box plot of relative bitrate error distribution across
the VCAs. The numbers displayed on the whiskers represent the mean relative absolute
error (MRAE). The error distribution and the MRAE values \rev{exhibhit}{exhibit} 
similar values for both \ipml and \rtpml methods across all three VCAs. For
example, in the case of \meet, the \ipml predictions are within 25\% of
theground truth bitrate
in 87\% of cases, while in \teams, it is 89\%, and in \webex, it is 95\%.
Comparatively, in the \rtpml method, these percentages are 89\%, 91\%, and 95\% for \meet, \teams,
and \webex, respectively.

We observe higher errors for both heuristics compared to the
ML methods, except in the case of \teams. Moreover, the errors are
systemic, with median relative bitrate error consistently exceeding zero
across all VCAs for both heuristics. This is because neither of these
heuristics considers any application-layer overheads, such as due to
encoding metadata. It should be noted that we do consider the
overhead due to the fixed portion of the RTP headers, i.e., 12 bytes.
However, incorporating encoding overheads remains challenging even with
RTP headers, as those parts of the traffic are encrypted. The ML methods,
on the other hand, can address these systemic errors by training on
video bitrate values observed at the application level.

\paragraph{Feature importance for \ipml method}. Figure~\ref{fig:
fimp-bitrate-webex-ip} shows the top 5 important features of the \ipml method
in the case of \webex. As expected, the feature \textit{\# bytes} has the
highest importance. In fact, that is the case across all
three VCAs. Most of the other important features
also relate to data volumes, such as \textit{Size [mean]} and
\textit{\# packets}. Interestingly, we do not observe any
semantics-based features among the top-5 features, except for \textit{\# unique
sizes}, which appears as the fourth most \rev{imortant}{important} feature for \webex. This
is because video bitrate is inherently correlated
with observed throughput. In fact, the top 5 features of the
\rtpml method are also found to be derived from flow statistics (see
Figure~\ref{fig:bitrate-fimp-rtp} \rev{}{ in Appendix}).


\subsubsection{Frame Jitter}

\begin{figure}[t!]
  \begin{center}
    \includegraphics[width=0.75\linewidth]{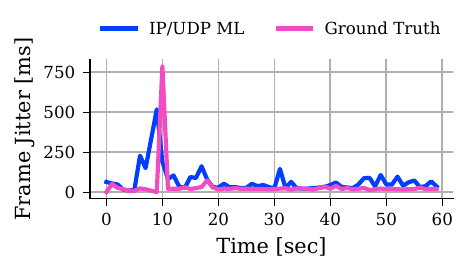}
  \end{center}
  \caption{A time series for frame jitter \ipml predictions over a single Meet trace}
  \label{fig:jitter_time_series}
\end{figure}

Figure~\ref{fig:jitter_error} shows the boxplot of the errors in frame
jitter predictions for the three VCAs. It is evident that all methods, including the
RTP-based approaches, tend to overestimate frame jitter in most cases.
Furthermore, we find that the MAE values are unusually high for this metric.
The average ground truth frame jitter observed across all three VCAs
falls within the range of 27-33 ms, which is comparable to the MAE
values obtained from all
methods. Upon further examination, we discover that the WebRTC ground-truth
statistic reports the jitter over decoded frames, encompassing
additional application delays such as jitter buffer and decoding delays.
The jitter buffer introduces variable delays to ensure
smooth video playback, while decoding delays vary based on the
client's computational resources. Capturing these variable
application-level delays can be challenging using only the network data.

Figure~\ref{fig:jitter_time_series} illustrates this
phenomenon with the frame jitter values reported
by the \ipml and WebRTC for a \meet call. The \ipml method
reports several spikes in
frame jitter throughout the call. While most of the smaller spikes seem to
be smoothed out in the WebRTC data, there is a significant spike around t=10s
that appears in both cases. The \ipml method estimates
spikes before t=10s, indicating significant jitter in frame arrival around
that time. The application jitter buffer might have attempted \rev{to}{}
to mitigate this frame jitter by emitting frames at a constant rate
until it is emptied, resulting in a larger spike later.

From a network
operator's perspective, it is more important to predict and respond to
network-level frame jitter. Ensuring a smooth frame arrival will
automatically lead to low frame jitter. In future work, we plan to modify our
experiment methods to collect ground truth frame jitter calculated
before the frame is enqueued to the jitter buffer. This will allow us to
assess the error of our method more accurately by providing a reliable
basis for comparison.

\begin{figure}
  \begin{center}
    \includegraphics[width=0.75\linewidth]
    {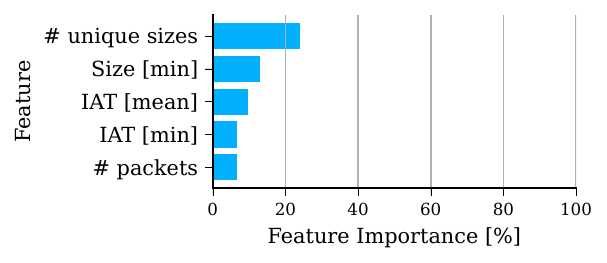}
  \end{center}
  \caption{Top-5 feature importance scores for \ipml resolution predictions for Webex}
  \label{fig:feat-webex-res-lab}
\end{figure}

\subsubsection{Resolution} \label{sec:lab-resolution}

\begin{table}[h]
  \centering
  \small
  \begin{tabular}{|c|ccc|}
    \hline
    \multirow{2}{*}{\textbf{Method}} & \multicolumn{3}{c|}{\textbf{Accuracy}}                                                        \\ \cline{2-4}
                                     & \multicolumn{1}{c|}{\textbf{Meet}}     & \multicolumn{1}{c|}{\textbf{Teams}} & \textbf{Webex} \\ \hline
    \ipml                            & \multicolumn{1}{c|}{97.74\%}           & \multicolumn{1}{c|}{87.22\%}        & 99.30\%        \\ \hline
    \rtpml                           & \multicolumn{1}{c|}{97.87\%}           & \multicolumn{1}{c|}{87.78\%}        & 99.31\%        \\ \hline
  \end{tabular}
  \caption{Resolution estimation accuracy across VCAs}
  \label{tab:resolution-acc}
\end{table}

\begin{table}[h]
  \centering
  \small
  \begin{tabular}{|c|ccc|c|}
    \hline
    \multirow{2}{*}{\textbf{Actual}} & \multicolumn{3}{c|}{\textbf{Predicted}} & \multirow{2}{*}{\textbf{Total}}                  \\ \cline{2-4}
                                     & \multicolumn{1}{c|}{Low}                & \multicolumn{1}{c|}{Medium}     & High    &      \\ \hline
    Low                              & \multicolumn{1}{c|}{96.41\%}            & \multicolumn{1}{c|}{1.65\%}     & 1.95\%  & 5038 \\ \hline
    Medium                           & \multicolumn{1}{c|}{8.08\%}             & \multicolumn{1}{c|}{45.40\%}    & 46.52\% & 1782 \\ \hline
    High                             & \multicolumn{1}{c|}{1.20\%}             & \multicolumn{1}{c|}{7.85\%}     & 90.95\% & 7588 \\ \hline
  \end{tabular}\caption{The normalized confusion matrix for resolution predictions
    by \ipml model for Teams.} 
  \label{tab:resolution-cm}
\end{table}

We use frame height as the measure for resolution. Within our dataset,
we observe 3 distinct frame height values for \meet:
180, 270, and 360; 11 distinct values for \teams ranging from 90 to
720; and only 2 distinct values for \webex: 180 and 360. For \meet and
\webex, we apply classification on a
per-value basis. For \teams, we bin the frame height into three classes: \textit{low} ($\leq$
240), \textit{medium} ((240, 480]), and \textit{high} ($>$ 480). Table \ref{tab:resolution-acc} shows the overall resolution accuracies across all VCAs. In all cases, the
accuracy is comparable to that of the \rtpml method.

Table~\ref{tab:resolution-cm} shows the confusion matrix for \teams
using the \ipml method. It is evident that the \ipml method
accurately predicts the \textit{low} and \textit{high} resolution
classes. However, it misclassifies 46.52\% of \textit{medium} resolution
intervals as \textit{high} resolution. This discrepancy could be
attributed to either class imbalance in one or more
of the 5-fold cross-validation splits or
the inherent difficulty in distinguishing between the \textit{medium}
and \textit{high}
resolution classes. Notably, within the \textit{medium}
resolution bin, 70\%
of the intervals have a frame height of
404, which is close to the threshold of 480 used to differentiate
\textit{medium} and
\textit{high} resolution classes.

\paragraph{Feature importance.} For the \ipml method, packet size
statistics consistently appear in the top 5 features for all VCAs. In fact, for \meet and \teams,
3 out of top 5 features are related to packet sizes, suggesting a
strong correlation between frame resolution and packet sizes. For \webex (see Figure~\ref{fig:feat-webex-res-lab}), the most important feature is \textit{\# unique sizes},
indicating a correlation between frame rate and
frame resolution. We find similar patterns in feature importance plots
for the \rtpml method (see Figure~\ref{fig:resolution-fimp-rtp} \rev{}{ in Appendix}). The only exception is
\webex, where the \textit{\# unique sizes} feature is replaced by \textit{$unique RTP_{vid} TS$} and \textit{$Marker_{vid}$ bit sum} features. This finding re-affirms that packet size difference is valuable for identifying frame boundaries.

\begin{figure*}[t!]
  \centering
  \begin{subfigure}{0.3\textwidth}
    \includegraphics[width=1.0\textwidth]
    {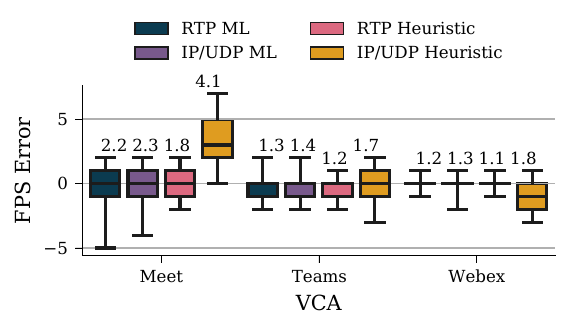}
    \caption{Frame rate} \label{subfig:fps-rw}
  \end{subfigure}
  \begin{subfigure}{0.3\textwidth}
    \centering
    \includegraphics[width=1.0\textwidth]
    {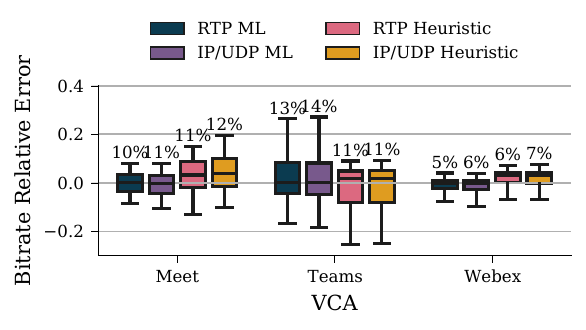}
    \caption{Bitrate}
    \label{subfig:bitrate-rw}
  \end{subfigure}
  \begin{subfigure}{0.3\textwidth} \centering
    \includegraphics[width=1.0\textwidth]
    {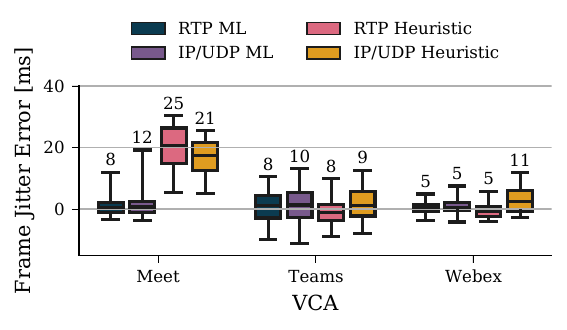}
    \caption{Frame jitter} \label{subfig:jitter-rw} \end{subfigure}
  \caption{Distribution of errors across the VCAs for the real-world
    dataset. The
    whiskers represent the $10^{th}$ and $90^{th}$ percentile values. The numbers
    above the top whisker represent the MAE values for frame rate and
    frame jitter and MRAE for bitrate.}
  \label{fig:qoe-rw} \end{figure*}

\subsection{Real-world data} \label{sec:realworld}
This section describes the results of the data collected from
\ndevices access networks. We observe a few differences in the
real-world dataset compared to the in-lab dataset. The
\texttt{payload type} in \teams and \webex is different: For \teams, we observe a \texttt{payload type} of 100
for video, 101 for video retransmission, while for \webex, the
\texttt{payload type} for video is 100, with no retransmissions as in the lab
data. We adjust the media classification approach for the RTP methods
accordingly, while the remaining methodology remains the same as for the in-lab data.

\subsubsection{Frame Rate}
Figure~\ref{subfig:fps-rw} shows the boxplot of frame rate
estimation errors. The overall accuracy is high for the \ipml method
and is comparable to the \rtpml method, a difference of 0.1~FPS
across all VCAs. Interestingly, the \rtpnonml has the highest accuracy
\rev{amongst}{among} all methods. We believe it could be because
network conditions are more stable in real-world data, thus
reducing any errors in \rtpnonml due to application-level delays
such as jitter buffer delay.

The \ipnonml, on the other hand, has the highest errors among all
methods. While the MAE difference between the \ipnonml and \rtpnonml is only 0.5~FPS and 0.7~FPS for
\teams and \webex, it is 2.3~FPS for \meet. Upon further inspection, we
find that the higher
errors for \meet are because of a higher fraction of frames in the
real-world data where the intra-frame packet size difference is greater
than the $\Delta_{size}^{max}$, the threshold used to determine frame
boundaries. More specifically, in the lab data, the intra-frame size
difference exceeded $\Delta_{size}^{max}$ for only 4.26\% of frames, while
this number is 14.48\% in the real-world data. This also explains
consistent overestimation for \meet. Note that using a higher value for
$\Delta_{size}^{max}$ will not help as it will lead to underestimation due to combining
frames with similar sizes. The discrepancy in \meet could be
a codec-specific issue; \meet uses VP8 or VP9, while both
\teams and \webex use H.264, leading to the fragmentation of frames into
unequal-sized packets. We will examine this further in our future work.

We also notice this anomaly in the \ipml feature importance analysis. While \textit{\# unique sizes} is \rev{amongst}{among} the top 5 features
 for \teams and \webex,  it is not the case for \meet. 
 Instead, this is replaced by
the IAT statistics, indicating that packet arrival patterns are better
signals for detecting frame
boundaries. This finding confirms that \textit{\# unique sizes} is not as strongly correlated with frame
rate for \meet in the real-world data. This also shows the resiliency
of ML models as they can rely on multiple features together more
effectively.

\subsubsection{Bitrate}
Figure~\ref{subfig:bitrate-rw} shows the boxplot of relative error
distribution with overall MRAE values mentioned over the top whisker.
The MRAE values in the real-world data are smaller than the
in-lab data across all methods. For example, the \ipml method can estimate bitrate within 25\% of
ground truth in 92.17\% of the intervals for Meet, 82.43\% for Teams, and 95.14\%
for Webex. This is likely because the bitrate
values are more stable, making predictions easier.
The feature importance trends for bitrate are similar to
in-lab data for each VCA. The most important features for both \rtpml
and \ipml are again derived from flow statistic and correspond to data
volume features such as \textit{\# bytes} and \textit{\# packets}.

\subsubsection{Frame Jitter}
We observe that the overall frame jitter errors are lower in the real-world
data than the in-lab data for most methods (see Figure~\ref{subfig:jitter-rw} and Figure~\ref{fig:jitter_error}). For example, when
analyzing the \ipml MAE value for \meet, the MAE is 9.3~ms for real-world
data, whereas it is 22.6~ms for in-lab data. This
difference is likely because the network conditions are more stable in
the real-world dataset. This leads to a lower network-level frame jitter,
reducing the smoothening effect of the application-level delay jitter
buffer. Thus, the differences
between the predicted frame jitter (only network-data) and the
WebRTC frame jitter (that includes effect of application delay jitter buffer) will be smaller,
leading to reduced overall errors. The remaining trends are similar to the
in-lab data.

\subsubsection{Resolution}
The real-world dataset for \meet contains two additional frame height
values: 540 and 720. This is likely because of greater throughput
availability and explains the greater overall bitrate values for
\meet. For \teams, the same set of resolution values were
observed as in-lab data. We only observe a single
resolution for \webex, and thus skip its accuracy computation.

The accuracy for resolution classification using \ipml is 96.26\% and
86.82\% for \meet and \teams, respectively. This is comparable to the \rtpml
accuracy -- 96.75\% for \meet and 87.11\% for \teams, respectively.
Similar to the in-lab
data, the \ipml model can distinguish extreme
resolution values (see Table~\ref{tab:resolution-cm-rw} for \teams) with
high accuracy, while the accuracy is low for \textit{medium}
resolution intervals.

\subsection{Model Transferability}

\begin{table}[t!]
  \centering
  \small
  \begin{tabular}{|c|ccc|}
    \hline
    \multirow{2}{*}{\textbf{Method}} & \multicolumn{3}{c|}{\textbf{VCA}}                                                         \\ \cline{2-4}
                                     & \multicolumn{1}{c|}{\textbf{Meet}} & \multicolumn{1}{c|}{\textbf{Teams}} & \textbf{Webex} \\ \hline
    \ipml                            & \multicolumn{1}{c|}{12.41}         & \multicolumn{1}{c|}{2.07}           & 1.56           \\ \hline
    \rtpml                           & \multicolumn{1}{c|}{3.11}          & \multicolumn{1}{c|}{2.51}           & 1.51           \\ \hline
  \end{tabular}
  \caption{Frame rate MAE results after using lab-trained models to predict
    real-world MAE}
  \label{tab:generalize}
\end{table}

We examine the transferability of ML models by testing the in-lab
trained ML models with the real-world dataset. Table~\ref{tab:generalize} shows the overall MAE values for frame
rate estimation. When considering the \ipml approach, MAE increases
marginally for both \teams and \webex, specifically by 0.7~FPS and 0.3~FPS,
respectively, compared to using models trained on real-world
data. However, for \meet, the MAE significantly increases by 10~FPS.
Upon further
inspection, we find that \textit{IAT [min]} is the most important
feature for the in-lab-trained \ipml model. Considering the
disparity in bitrates between real-world and lab data for \meet, it is likely
that the IAT distribution also differs, leading to
errors in frame rate prediction. Interestingly, the decline in performance for \meet using the \rtpml
method is not as pronounced as observed in the \ipml method. This
disparity can be attributed to the higher importance of the number of
unique RTP timestamps as a feature, which in some sense is a more direct
indicator of frame rate compared to IAT.

The trend persists for video bitrate and resolution, with a significant drop
in accuracy for \meet but only a slight decrease for \teams and \webex (see
Tables~\ref{tab:bitrate-transfer} and \ref{tab:jitter-transfer} in the Appendix).
The non-transferability for \meet can again be attributed to the presence of
a distinct distribution that was not previously encountered, i.e., calls
with high bitrate and high
resolution. This discrepancy suggests that the model cannot effectively extrapolate to unseen distributions.

\subsection{Effect of Network Conditions}
\begin{figure}[t!]
  \centering
  \includegraphics[width=0.75\linewidth]
  {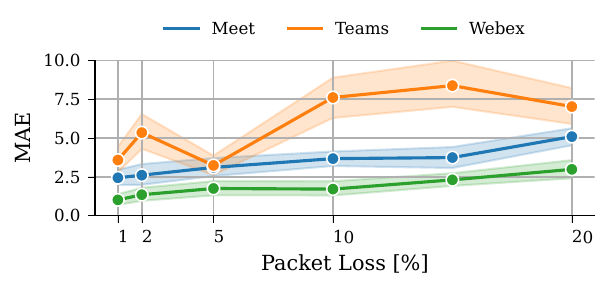}
  \caption{\ipml MAE for frame rate with varying network loss. The bands
    represent $95\%$ confidence intervals}
  \label{fig:fr-loss}
\end{figure}

We next characterize the network conditions under which the models yield
high errors. To do so, we collect data under \rev{synethetic}{synthetic} network conditions by
varying one of the following five network parameters: throughput
(1500~kbps),
throughput jitter (0~kbps), latency (50~ms), latency jitter (0~ms), and
packet loss (0\%). The numbers in parentheses represent the
default values. For example, to analyze the impact of loss,
other parameters are set to default values, and loss is varied from
0\% to 20\% following a Bernoulli loss model. Each
combination of network conditions is repeated for four calls. For
training ML
models, we use 50\% of data, sampling uniformly randomly from each
combination of network condition.  The remaining 50\% data is used for
testing.

Figure~\ref{fig:fr-loss} shows the accuracy under varying loss for the \ipml method. Barring few exceptions, we observe an increasing trend in
errors as network loss increases. On further inspection, we found that
losses lead to retransmissions for video packets, leading to
packet reordering. It is not possible to determine the correct order of the packets
using only IP/UDP headers which causes higher errors. We find that the
errors are even higher for the \ipnonml as it relies only on packet
sizes, and is more severely impacted by packet reordering. We also
observe similar behavior under high latency or throughput jitter likely
because both also lead to
packet reordering. However, this occurs at very high values of jitter,
indicating some robustness to minor jitters in the network. The errors do not change significantly with varying
mean throughput or mean latency.

\subsection{Effect of Prediction Window Size}

\begin{figure}[t!]
  \begin{center}
    \includegraphics[width=0.75\linewidth]{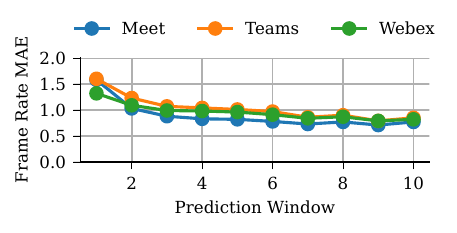}
  \end{center}
  \caption{Variation of \ipml MAE with prediction window size for frame rate
    predictions for in-lab traces}
  \label{fig:window_size_effect}
\end{figure}

We analyze the impact of prediction window size on QoE estimation accuracy.
Figure~\ref{fig:window_size_effect} shows the \ipml MAE values for frame rate
under varying prediction window. The errors decrease
as the prediction window size increases. This can be attributed to two reasons:
(1). larger window sizes reduce the impact of sub-second-level window misalignment between packet traces
and WebRTC logs, and (2). the frame rate values
become more stable as they are smoothed out over larger window, making
the prediction task easier. We observe similar patterns across other
methods and metrics.

  \section{Related Work}\label{sec:related}
\paragraph{QoE Inference for Video Streaming}. Past research has made
substantial progress in inferring QoE for
on-demand video streaming. One set
of approaches propose heuristics that model a video session
relying on the properties of the underlying streaming protocol~\cite{Dimopoulos2013, Schatz2012, mangla2018emimic}.
The second set of approaches propose using supervised machine learning
and use features derived from network data to estimate QoE metrics~\cite{Mazhar2018,
    bronzino2019inferring, BUFFEST, Aggarwal2014}. Inferring QoE for video
conferencing is a fairly distinct problem from video streaming due to the differences in the
nature of two applications, consequently leading to differences in the underlying application and transport protocols,
and the metrics that determine user QoE. This paper tackles the problem
of QoE inference for VCAs and proposes both heuristic- and
ML-based approaches.

\textbf{VCA measurement studies.} Early VCA measurement studies focused  on understanding the design and
network performance of Skype, one of the first and the most popular VCA
of the time~\cite{baset2004analysis, guha2005experimental, hossfeld2008analysis}.
More recent studies have revisited similar questions for modern
VCAs~\cite{macmillan2021measuring, nistico2020comparative,
    jansen2018performance, chang2021can, he2023measurement}. Most of these studies rely on controlled
experiments and assume access to end-hosts to collect VCA performance data. For instance, He et
al.~\cite{he2023measurement} identify the functional differences (e.g.,
congestion control mechanisms) among modern VCAs using controlled
measurements. Our work considers a different question, i.e., how to
infer video QoE metrics without access to end-hosts? Answering this
question can enable network operators to
understand VCA performance for a wide-variety of application
and network contexts and appropriately manage their networks.

\textbf{VCA QoE inference.}  Past work has proposed data-driven techniques, based on supervised machine learning, to estimate QoE for
\textit{Voice over IP}~\cite{chen2006quantifying, Aggarwal2014}. More recent works propose similar techniques but focus on \textit{video} performance over VCAs. \rev{More recent works
propose similar techniques to infer video performance among
VCAs. In doing so, these works assume access to
RTP headers which may not be practical in many cases such as with custom RTP protocols (e.g., Zoom), encrypted application-layer headers (e.g., VPN), or legacy monitoring systems.}{These works differ, however, in the set of inferred QoE metrics as well as the network features used for inference. For instance, Garcia et al. infer metrics assuming access to an unimpaired reference video\cite{garcia2020assessment}. Similarly,  Yan et al.\cite{Yan2015EnablingQL} use WiFi-specific features to predict ``good versus bad'' QoE over the entire VCA session. We focus on inferring no-reference, \textit{objective} VCA QoE metrics using measurements of passive network traffic. Works by Nikravesh et al.\cite{nikravesh2016qoe} and Carofigilo et al.~\citep{carofiglio2021characterizing} are similar in spirit in that regard. However, both of these {works} assume access to
RTP headers which may not be practical in many cases such as with custom RTP
protocols (e.g., Zoom), encrypted application-layer headers (e.g., VPN), or legacy
monitoring systems.} Recent work by Oliver et al.~\cite{zoom_rtp}
uses entropy-based header analysis to infer Zoom's RTP encapsulation
mechanisms. However, the approach may not work if VCAs use complex
encapsulation mechanisms or encrypt application-layer headers
altogether. Morevoer, it requires network monitoring systems
that can process arbitrary portions of the traffic. This may not be
feasible for several network operators due to practical considerations.
This paper considers whether more standard features of the network
traffic, i.e., IP/UDP headers, can be used to infer the VCA QoE metrics.

  \section{Limitations and Future Work}
\label{sec:discussion}

\paragraph{Generalizability to other VCAs.} Our paper's evaluation
is focused on WebRTC-based VCAs, although our methodology can be applied
to any RTP-based VCA. The reason to
focus on WebRTC is the lack of methods to obtain application-level QoE
metrics for native VCA clients. Additionally, we do not include the WebRTC
version of Zoom, one of the most popular VCAs, as its implementation  
uses the \texttt{datachannel} API meant for
non-audiovisual communication.  As a result, the video
QoE metrics are no longer available for Zoom through the \texttt{webrtc-internals} API.
Past work has considered other metrics to obtain QoE metrics from the
applications. Michel et
al.~\cite{michel2022enabling} used a custom Zoom client, but this
approach will not work for the native client of other VCAs. 
Another method
to obtain application-level logs is through screen
capture of annotated video~\cite{salsify, varvello2022performance}, but this
method is resource intensive. Future work
will explore generalizable and lightweight methods to obtain
application-level QoE logs for native VCA clients and assess the
accuracy of
proposed QoE estimation methods for these clients.  

\paragraph{Cost of ML models}. Using supervised ML models can be costly due to the expense of acquiring labeled data for training.
We present one solution to gather labeled data, i.e., through automated data collection
frameworks, deployed either in-lab or across multiple network vantage
points. The framework is easily
extensible to other WebRTC-based VCAs. Another solution to explore in
future would be whether
direct or calibrated estimations from non-machine learning methods like
\ipnonml or \rtpnonml can be used as alternatives to labeled data.

\paragraph{Impact of application modes}. We only evaluate our
methodology in a two-person call scenario. However, modern VCAs offer
various other application modes, such as disabling video, multi-party
conferencing, and screen sharing. Determining whether user video is
disabled seems possible by analyzing UDP packet size distribution, but
the other two  modes pose challenges in QoE estimation, especially
using only IP/UDP headers. In multi-party scenarios,
multiple video streams may be transmitted over the same UDP flow. This may require
an additional step in our methods to estimate the number of
participants before estimating QoE. Similarly, when
screen sharing is enabled, adjustments to the media
classification steps will be required. These adjustments may be based on insights from differences in
encoding of video and screen sharing data.  Additionally, a machine learning-based QoE
inference approach such as \ipml, when trained with appropriate data,
could accurately estimate QoE metrics even across different application modes.
Further research will explore this question and quantify the impact of
application modes on the accuracy of our methods.

\paragraph{System considerations}. In theory, our approach relies on
lightweight features from the IP/UDP headers of network traffic.
However, we have not tested the scalability of our methods on a network-wide
level, particularly when it comes to real-time QoE estimation.
Additional optimization might be required in the implementation of our
methods such as using efficient data structures or implementation of
streaming versions of the methods.  In future work, we 
 plan to implement these approaches within a real-world network, 
such as campus network, to assess the scalability of our approach. 

  \section{Conclusion}\label{sec:conclusion}

We have developed and evaluated two methods to infer QoE for WebRTC-based VCAs at
per-second granularity. Evaluation of our method under diverse network
conditions demonstrates the model's ability to estimate QoE metrics with high
accuracy, even if the methods relies on only IP/UDP headers. This approach
represents a significant advance over previous work, which uses
information in the RTP headers. Future work will explore the
generalizability of our methods to a broader set of clients (e.g., device,
operating systems, native clients) and
application modes (e.g., multi-party calls).

  \section*{Acknowledgements}

We would like to thank our shepherd, Gianni Antichi and the anonymous reviewers for their valuable feedback. We also thank Guilherme Martins for helping with the real-world data collection and Irene
Pattarachanyakul for her efforts towards earlier version of this research. The work was supported by National Science Foundation (NSF) awards CNS-2124393, CNS-2223610, CNS-2224687, OAC-2126327, CNS-2146496, CNS-2131826, and CNS-2313190 at the University of Chicago, and Intel and NSF Awards CNS-2003257, OAC-2126327, and OAC-2126281 at UCSB.

\microtypesetup{protrusion=false}

\bibliographystyle{ACM-Reference-Format}
\thispagestyle{empty}
\balance \bibliography{paper}


\begin{thebibliography}{45}


\ifx \showCODEN    \undefined \def \showCODEN     #1{\unskip}     \fi
\ifx \showDOI      \undefined \def \showDOI       #1{#1}\fi
\ifx \showISBNx    \undefined \def \showISBNx     #1{\unskip}     \fi
\ifx \showISBNxiii \undefined \def \showISBNxiii  #1{\unskip}     \fi
\ifx \showISSN     \undefined \def \showISSN      #1{\unskip}     \fi
\ifx \showLCCN     \undefined \def \showLCCN      #1{\unskip}     \fi
\ifx \shownote     \undefined \def \shownote      #1{#1}          \fi
\ifx \showarticletitle \undefined \def \showarticletitle #1{#1}   \fi
\ifx \showURL      \undefined \def \showURL       {\relax}        \fi
\providecommand\bibfield[2]{#2}
\providecommand\bibinfo[2]{#2}
\providecommand\natexlab[1]{#1}
\providecommand\showeprint[2][]{arXiv:#2}

\bibitem[tcp(2022)]%
        {tcp-info-mlab}
 \bibinfo{year}{2022}\natexlab{}.
\newblock \bibinfo{title}{The M-Lab TCP INFO Data Set}.
\newblock
\newblock
\urldef\tempurl%
\url{https://measurementlab.net/tests/tcp-info}
\showURL{%
\tempurl}


\bibitem[mla(2022)]%
        {mlab-data}
 \bibinfo{year}{2022}\natexlab{}.
\newblock \bibinfo{title}{{Measurement Lab -- Network Diagnostic Test (NDT7)}}.
\newblock
\newblock
\urldef\tempurl%
\url{https://www.measurementlab.net/tests/ndt/ndt7/}
\showURL{%
\tempurl}


\bibitem[web(2023)]%
        {webrtc-internals}
 \bibinfo{year}{2023}\natexlab{}.
\newblock \bibinfo{booktitle}{\emph{Chrome WebRTC Internals}}.
\newblock
\urldef\tempurl%
\url{chrome://webrtc-internals}
\showURL{%
\tempurl}


\bibitem[Aggarwal et~al\mbox{.}(2014)]%
        {Aggarwal2014}
\bibfield{author}{\bibinfo{person}{Vaneet Aggarwal}, \bibinfo{person}{Emir
  Halepovic}, \bibinfo{person}{Jeffrey Pang}, \bibinfo{person}{Shobha
  Venkataraman}, {and} \bibinfo{person}{He Yan}.}
  \bibinfo{year}{2014}\natexlab{}.
\newblock \showarticletitle{Prometheus: Toward Quality-of-experience Estimation
  for Mobile Apps from Passive Network Measurements}. In
  \bibinfo{booktitle}{\emph{Proc. of ACM HotMobile}}.
\newblock


\bibitem[Banitalebi-Dehkordi et~al\mbox{.}(2015)]%
        {banitalebi2015effect}
\bibfield{author}{\bibinfo{person}{Amin Banitalebi-Dehkordi},
  \bibinfo{person}{Mahsa~T Pourazad}, {and} \bibinfo{person}{Panos
  Nasiopoulos}.} \bibinfo{year}{2015}\natexlab{}.
\newblock \showarticletitle{The effect of frame rate on 3D video quality and
  bitrate}.
\newblock \bibinfo{journal}{\emph{3D Research}} \bibinfo{volume}{6},
  \bibinfo{number}{1} (\bibinfo{year}{2015}), \bibinfo{pages}{1--13}.
\newblock


\bibitem[Baset and Schulzrinne(2004)]%
        {baset2004analysis}
\bibfield{author}{\bibinfo{person}{Salman~A Baset} {and}
  \bibinfo{person}{Henning Schulzrinne}.} \bibinfo{year}{2004}\natexlab{}.
\newblock \showarticletitle{An analysis of the skype peer-to-peer internet
  telephony protocol}.
\newblock \bibinfo{journal}{\emph{arXiv preprint cs/0412017}}
  (\bibinfo{year}{2004}).
\newblock


\bibitem[Borgolte et~al\mbox{.}(2019)]%
        {borgolte2019dns}
\bibfield{author}{\bibinfo{person}{Kevin Borgolte}, \bibinfo{person}{Tithi
  Chattopadhyay}, \bibinfo{person}{Nick Feamster}, \bibinfo{person}{Mihir
  Kshirsagar}, \bibinfo{person}{Jordan Holland}, \bibinfo{person}{Austin
  Hounsel}, {and} \bibinfo{person}{Paul Schmitt}.}
  \bibinfo{year}{2019}\natexlab{}.
\newblock \showarticletitle{How dns over https is reshaping privacy,
  performance, and policy in the internet ecosystem}. In
  \bibinfo{booktitle}{\emph{TPRC47: The 47th Research Conference on
  Communication, Information and Internet Policy}}.
\newblock


\bibitem[Bronzino et~al\mbox{.}(2019)]%
        {bronzino2019inferring}
\bibfield{author}{\bibinfo{person}{Francesco Bronzino}, \bibinfo{person}{Paul
  Schmitt}, \bibinfo{person}{Sara Ayoubi}, \bibinfo{person}{Guilherme Martins},
  \bibinfo{person}{Renata Teixeira}, {and} \bibinfo{person}{Nick Feamster}.}
  \bibinfo{year}{2019}\natexlab{}.
\newblock \showarticletitle{Inferring streaming video quality from encrypted
  traffic: Practical models and deployment experience}.
\newblock \bibinfo{journal}{\emph{Proceedings of the ACM on Measurement and
  Analysis of Computing Systems}} \bibinfo{volume}{3}, \bibinfo{number}{3}
  (\bibinfo{year}{2019}), \bibinfo{pages}{1--25}.
\newblock


\bibitem[Carofiglio et~al\mbox{.}(2021)]%
        {carofiglio2021characterizing}
\bibfield{author}{\bibinfo{person}{Giovanna Carofiglio},
  \bibinfo{person}{Giulio Grassi}, \bibinfo{person}{Enrico Loparco},
  \bibinfo{person}{Luca Muscariello}, \bibinfo{person}{Michele Papalini}, {and}
  \bibinfo{person}{Jacques Samain}.} \bibinfo{year}{2021}\natexlab{}.
\newblock \showarticletitle{Characterizing the Relationship Between Application
  QoE and Network QoS for Real-Time Services}. In
  \bibinfo{booktitle}{\emph{Proceedings of the ACM SIGCOMM 2021 Workshop on
  Network-Application Integration}}.
\newblock


\bibitem[Chai et~al\mbox{.}(2019)]%
        {chai2019importance}
\bibfield{author}{\bibinfo{person}{Zimo Chai}, \bibinfo{person}{Amirhossein
  Ghafari}, {and} \bibinfo{person}{Amir Houmansadr}.}
  \bibinfo{year}{2019}\natexlab{}.
\newblock \showarticletitle{On the Importance of Encrypted-SNI (ESNI)) to
  Censorship Circumvention}. In \bibinfo{booktitle}{\emph{9th USENIX Workshop
  on Free and Open Communications on the Internet (FOCI 19)}}.
\newblock


\bibitem[Chang et~al\mbox{.}(2021)]%
        {chang2021can}
\bibfield{author}{\bibinfo{person}{Hyunseok Chang}, \bibinfo{person}{Matteo
  Varvello}, \bibinfo{person}{Fang Hao}, {and} \bibinfo{person}{Sarit
  Mukherjee}.} \bibinfo{year}{2021}\natexlab{}.
\newblock \showarticletitle{Can you see me now? A measurement study of Zoom,
  Webex, and Meet}. In \bibinfo{booktitle}{\emph{Proceedings of the 21st ACM
  Internet Measurement Conference}}. \bibinfo{pages}{216--228}.
\newblock


\bibitem[Chen et~al\mbox{.}(2006)]%
        {chen2006quantifying}
\bibfield{author}{\bibinfo{person}{Kuan-Ta Chen}, \bibinfo{person}{Chun-Ying
  Huang}, \bibinfo{person}{Polly Huang}, {and} \bibinfo{person}{Chin-Laung
  Lei}.} \bibinfo{year}{2006}\natexlab{}.
\newblock \showarticletitle{Quantifying {Skype} user satisfaction}. In
  \bibinfo{booktitle}{\emph{Proceedings of the 2006 conference on Applications,
  technologies, architectures, and protocols for computer communications}}
  (Pisa, Italy) \emph{(\bibinfo{series}{SIGCOMM '06})}.
  \bibinfo{publisher}{ACM}, \bibinfo{pages}{399--410}.
\newblock
\showISBNx{1-59593-308-5}
\urldef\tempurl%
\url{https://doi.org/10.1145/1159913.1159959}
\showDOI{\tempurl}


\bibitem[Dimopoulos et~al\mbox{.}(2013)]%
        {Dimopoulos2013}
\bibfield{author}{\bibinfo{person}{G. Dimopoulos}, \bibinfo{person}{P.
  Barlet-Ros}, {and} \bibinfo{person}{J. Sanjuas-Cuxart}.}
  \bibinfo{year}{2013}\natexlab{}.
\newblock \showarticletitle{{Analysis of YouTube user experience from passive
  measurements}}. In \bibinfo{booktitle}{\emph{Proc.~CNSM}}.
\newblock


\bibitem[Dimopoulos et~al\mbox{.}(2016)]%
        {dimopoulos2016measuring}
\bibfield{author}{\bibinfo{person}{Giorgos Dimopoulos}, \bibinfo{person}{Ilias
  Leontiadis}, \bibinfo{person}{Pere Barlet-Ros}, {and}
  \bibinfo{person}{Konstantina Papagiannaki}.} \bibinfo{year}{2016}\natexlab{}.
\newblock \showarticletitle{Measuring video QoE from encrypted traffic}. In
  \bibinfo{booktitle}{\emph{Proceedings of the 2016 Internet Measurement
  Conference}}. \bibinfo{pages}{513--526}.
\newblock


\bibitem[Farnaaz and Jabbar(2016)]%
        {farnaaz2016random}
\bibfield{author}{\bibinfo{person}{Nabila Farnaaz} {and} \bibinfo{person}{MA
  Jabbar}.} \bibinfo{year}{2016}\natexlab{}.
\newblock \showarticletitle{Random forest modeling for network intrusion
  detection system}.
\newblock \bibinfo{journal}{\emph{Procedia Computer Science}}
  \bibinfo{volume}{89} (\bibinfo{year}{2016}), \bibinfo{pages}{213--217}.
\newblock


\bibitem[Fouladi et~al\mbox{.}(2018)]%
        {salsify}
\bibfield{author}{\bibinfo{person}{Sadjad Fouladi}, \bibinfo{person}{John
  Emmons}, \bibinfo{person}{Emre Orbay}, \bibinfo{person}{Catherine Wu},
  \bibinfo{person}{Riad~S. Wahby}, {and} \bibinfo{person}{Keith Winstein}.}
  \bibinfo{year}{2018}\natexlab{}.
\newblock \showarticletitle{Salsify: {Low-Latency} Network Video through
  Tighter Integration between a Video Codec and a Transport Protocol}. In
  \bibinfo{booktitle}{\emph{15th USENIX Symposium on Networked Systems Design
  and Implementation (NSDI 18)}}. \bibinfo{publisher}{USENIX Association},
  \bibinfo{address}{Renton, WA}, \bibinfo{pages}{267--282}.
\newblock
\showISBNx{978-1-939133-01-4}
\urldef\tempurl%
\url{https://www.usenix.org/conference/nsdi18/presentation/fouladi}
\showURL{%
\tempurl}


\bibitem[Garcia et~al\mbox{.}(2019)]%
        {garcia2019understanding}
\bibfield{author}{\bibinfo{person}{Boni Garcia}, \bibinfo{person}{Micael
  Gallego}, \bibinfo{person}{Francisco Gortazar}, {and}
  \bibinfo{person}{Antonia Bertolino}.} \bibinfo{year}{2019}\natexlab{}.
\newblock \showarticletitle{Understanding and estimating quality of experience
  in WebRTC applications}.
\newblock \bibinfo{journal}{\emph{Computing}} \bibinfo{volume}{101},
  \bibinfo{number}{11} (\bibinfo{year}{2019}), \bibinfo{pages}{1585--1607}.
\newblock


\bibitem[Garc{\'\i}a et~al\mbox{.}(2020)]%
        {garcia2020assessment}
\bibfield{author}{\bibinfo{person}{Boni Garc{\'\i}a},
  \bibinfo{person}{Francisco Gort{\'a}zar}, \bibinfo{person}{Micael Gallego},
  {and} \bibinfo{person}{Andrew Hines}.} \bibinfo{year}{2020}\natexlab{}.
\newblock \showarticletitle{Assessment of qoe for video and audio in webrtc
  applications using full-reference models}.
\newblock \bibinfo{journal}{\emph{Electronics}} \bibinfo{volume}{9},
  \bibinfo{number}{3} (\bibinfo{year}{2020}), \bibinfo{pages}{462}.
\newblock


\bibitem[Guha and Daswani(2005)]%
        {guha2005experimental}
\bibfield{author}{\bibinfo{person}{Saikat Guha} {and} \bibinfo{person}{Neil
  Daswani}.} \bibinfo{year}{2005}\natexlab{}.
\newblock \bibinfo{booktitle}{\emph{An experimental study of the skype
  peer-to-peer voip system}}.
\newblock \bibinfo{type}{{T}echnical {R}eport}. \bibinfo{institution}{Cornell
  University}.
\newblock


\bibitem[He et~al\mbox{.}(2023)]%
        {he2023measurement}
\bibfield{author}{\bibinfo{person}{Jia He}, \bibinfo{person}{Mostafa Ammar},
  {and} \bibinfo{person}{Ellen Zegura}.} \bibinfo{year}{2023}\natexlab{}.
\newblock \showarticletitle{A Measurement-Derived Functional Model for the
  Interaction Between Congestion Control and QoE in Video Conferencing}. In
  \bibinfo{booktitle}{\emph{Passive and Active Measurement: 24th International
  Conference, PAM 2023, Virtual Event, March 21--23, 2023, Proceedings}}.
  Springer, \bibinfo{pages}{129--159}.
\newblock


\bibitem[Ho{\ss}feld and Binzenh{\"o}fer(2008)]%
        {hossfeld2008analysis}
\bibfield{author}{\bibinfo{person}{Tobias Ho{\ss}feld} {and}
  \bibinfo{person}{Andreas Binzenh{\"o}fer}.} \bibinfo{year}{2008}\natexlab{}.
\newblock \showarticletitle{Analysis of Skype VoIP traffic in UMTS: End-to-end
  QoS and QoE measurements}.
\newblock \bibinfo{journal}{\emph{Computer Networks}} (\bibinfo{year}{2008}).
\newblock


\bibitem[Hossfeld et~al\mbox{.}(2016)]%
        {hossfeld2016formal}
\bibfield{author}{\bibinfo{person}{Tobias Hossfeld}, \bibinfo{person}{Poul~E
  Heegaard}, \bibinfo{person}{Martin Varela}, {and} \bibinfo{person}{Sebastian
  M{\"o}ller}.} \bibinfo{year}{2016}\natexlab{}.
\newblock \showarticletitle{Formal definition of QoE metrics}.
\newblock \bibinfo{journal}{\emph{arXiv preprint arXiv:1607.00321}}
  (\bibinfo{year}{2016}).
\newblock


\bibitem[Huitema(2003)]%
        {huitema2003rfc3605}
\bibfield{author}{\bibinfo{person}{C Huitema}.}
  \bibinfo{year}{2003}\natexlab{}.
\newblock \bibinfo{title}{RFC3605: Real Time Control Protocol (RTCP) Attribute
  in Session Description Protocol (SDP)}.
\newblock
\newblock


\bibitem[Jansen et~al\mbox{.}(2018)]%
        {jansen2018performance}
\bibfield{author}{\bibinfo{person}{Bart Jansen}, \bibinfo{person}{Timothy
  Goodwin}, \bibinfo{person}{Varun Gupta}, \bibinfo{person}{Fernando Kuipers},
  {and} \bibinfo{person}{Gil Zussman}.} \bibinfo{year}{2018}\natexlab{}.
\newblock \showarticletitle{Performance evaluation of WebRTC-based video
  conferencing}.
\newblock \bibinfo{journal}{\emph{ACM SIGMETRICS Performance Evaluation
  Review}} \bibinfo{volume}{45}, \bibinfo{number}{3} (\bibinfo{year}{2018}),
  \bibinfo{pages}{56--68}.
\newblock


\bibitem[Jesup(2011)]%
        {rfc6184}
\bibfield{author}{\bibinfo{person}{R Jesup}.} \bibinfo{year}{2011}\natexlab{}.
\newblock \showarticletitle{RTP Payload Format for H. 264 Video}.
\newblock \bibinfo{journal}{\emph{Internet Eng. Task Force, RFC}}
  \bibinfo{volume}{6184} (\bibinfo{year}{2011}).
\newblock


\bibitem[Krishnamoorthi et~al\mbox{.}(2017)]%
        {BUFFEST}
\bibfield{author}{\bibinfo{person}{Vengatanathan Krishnamoorthi},
  \bibinfo{person}{Niklas Carlsson}, \bibinfo{person}{Emir Halepovic}, {and}
  \bibinfo{person}{Eric Petajan}.} \bibinfo{year}{2017}\natexlab{}.
\newblock \showarticletitle{{BUFFEST}: Predicting Buffer Conditions and
  Real-time Requirements of {HTTP(S)} Adaptive Streaming Clients}. In
  \bibinfo{booktitle}{\emph{Proc. of ACM MMSys}}.
\newblock


\bibitem[Li(2007)]%
        {rfc5109}
\bibfield{author}{\bibinfo{person}{A Li}.} \bibinfo{year}{2007}\natexlab{}.
\newblock \showarticletitle{RTP Payload Format for Generic Forward Error
  Correction, RFC 5109}.
\newblock  (\bibinfo{year}{2007}).
\newblock


\bibitem[MacMillan et~al\mbox{.}(2021)]%
        {macmillan2021measuring}
\bibfield{author}{\bibinfo{person}{Kyle MacMillan}, \bibinfo{person}{Tarun
  Mangla}, \bibinfo{person}{James Saxon}, {and} \bibinfo{person}{Nick
  Feamster}.} \bibinfo{year}{2021}\natexlab{}.
\newblock \showarticletitle{Measuring the performance and network utilization
  of popular video conferencing applications}. In
  \bibinfo{booktitle}{\emph{Proceedings of the 21st ACM Internet Measurement
  Conference}}.
\newblock


\bibitem[MacMillan et~al\mbox{.}(2023)]%
        {macmillan2023comparative}
\bibfield{author}{\bibinfo{person}{Kyle MacMillan}, \bibinfo{person}{Tarun
  Mangla}, \bibinfo{person}{James Saxon}, \bibinfo{person}{Nicole~P Marwell},
  {and} \bibinfo{person}{Nick Feamster}.} \bibinfo{year}{2023}\natexlab{}.
\newblock \showarticletitle{A Comparative Analysis of Ookla Speedtest and
  Measurement Labs Network Diagnostic Test (NDT7)}.
\newblock \bibinfo{journal}{\emph{Proceedings of the ACM on Measurement and
  Analysis of Computing Systems}} \bibinfo{volume}{7}, \bibinfo{number}{1}
  (\bibinfo{year}{2023}), \bibinfo{pages}{1--26}.
\newblock


\bibitem[Mangla et~al\mbox{.}(2018)]%
        {mangla2018emimic}
\bibfield{author}{\bibinfo{person}{Tarun Mangla}, \bibinfo{person}{Emir
  Halepovic}, \bibinfo{person}{Mostafa Ammar}, {and} \bibinfo{person}{Ellen
  Zegura}.} \bibinfo{year}{2018}\natexlab{}.
\newblock \showarticletitle{{eMIMIC: Estimating HTTP-based Video QoE Metrics
  from Encrypted Network Traffic}}. In \bibinfo{booktitle}{\emph{Proc. of
  IEEE/IFIP TMA}}.
\newblock


\bibitem[Mangla et~al\mbox{.}(2020)]%
        {mangla2020drop}
\bibfield{author}{\bibinfo{person}{Tarun Mangla}, \bibinfo{person}{Emir
  Halepovic}, \bibinfo{person}{Ellen Zegura}, {and} \bibinfo{person}{Mostafa
  Ammar}.} \bibinfo{year}{2020}\natexlab{}.
\newblock \showarticletitle{Drop the packets: using coarse-grained data to
  detect video performance issues}. In \bibinfo{booktitle}{\emph{Proceedings of
  the 16th International Conference on emerging Networking EXperiments and
  Technologies}}. \bibinfo{pages}{71--77}.
\newblock


\bibitem[Marczak and Scott-Railton(2021)]%
        {zoom_rtp}
\bibfield{author}{\bibinfo{person}{Bill Marczak} {and} \bibinfo{person}{John
  Scott-Railton}.} \bibinfo{year}{2021}\natexlab{}.
\newblock \bibinfo{title}{Move Fast and Roll Your Own Crypto: A Quick Look at
  the Confidentiality of Zoom Meetings}.
\newblock
\newblock
\urldef\tempurl%
\url{https://citizenlab.ca/2020/04/move-fast-roll-your-own-crypto-a-quick-look-at-the-confidentiality-of-zoom-meetings/}
\showURL{%
\tempurl}


\bibitem[Mazhar and Shafiq(2018)]%
        {Mazhar2018}
\bibfield{author}{\bibinfo{person}{M.~H. Mazhar} {and} \bibinfo{person}{Z.
  Shafiq}.} \bibinfo{year}{2018}\natexlab{}.
\newblock \showarticletitle{{Real-time Video Quality of Experience Monitoring
  for HTTPS and QUIC}}. In \bibinfo{booktitle}{\emph{Proc.~of IEEE INFOCOM}}.
\newblock


\bibitem[Michel et~al\mbox{.}(2022)]%
        {michel2022enabling}
\bibfield{author}{\bibinfo{person}{Oliver Michel}, \bibinfo{person}{Satadal
  Sengupta}, \bibinfo{person}{Hyojoon Kim}, \bibinfo{person}{Ravi Netravali},
  {and} \bibinfo{person}{Jennifer Rexford}.} \bibinfo{year}{2022}\natexlab{}.
\newblock \showarticletitle{Enabling passive measurement of zoom performance in
  production networks}. In \bibinfo{booktitle}{\emph{Proceedings of the 22nd
  ACM Internet Measurement Conference}}. \bibinfo{pages}{244--260}.
\newblock


\bibitem[Nikravesh et~al\mbox{.}(2016)]%
        {nikravesh2016qoe}
\bibfield{author}{\bibinfo{person}{Ashkan Nikravesh}, \bibinfo{person}{David~Ke
  Hong}, \bibinfo{person}{Qi~Alfred Chen}, \bibinfo{person}{Harsha~V
  Madhyastha}, {and} \bibinfo{person}{Z~Morley Mao}.}
  \bibinfo{year}{2016}\natexlab{}.
\newblock \showarticletitle{QoE inference without application control}. In
  \bibinfo{booktitle}{\emph{Proceedings of the 2016 workshop on QoE-based
  Analysis and Management of Data Communication Networks}}.
  \bibinfo{pages}{19--24}.
\newblock


\bibitem[Nistico et~al\mbox{.}(2020)]%
        {nistico2020comparative}
\bibfield{author}{\bibinfo{person}{Antonio Nistico}, \bibinfo{person}{Dena
  Markudova}, \bibinfo{person}{Martino Trevisan}, \bibinfo{person}{Michela
  Meo}, {and} \bibinfo{person}{Giovanna Carofiglio}.}
  \bibinfo{year}{2020}\natexlab{}.
\newblock \showarticletitle{A comparative study of RTC applications}. In
  \bibinfo{booktitle}{\emph{2020 IEEE International Symposium on Multimedia
  (ISM)}}. \bibinfo{pages}{1--8}.
\newblock


\bibitem[Schatz et~al\mbox{.}(2012)]%
        {Schatz2012}
\bibfield{author}{\bibinfo{person}{R. Schatz}, \bibinfo{person}{T. Hossfeld},
  {and} \bibinfo{person}{P. Casas}.} \bibinfo{year}{2012}\natexlab{}.
\newblock \showarticletitle{Passive {YouTube QoE} Monitoring for {ISPs}}. In
  \bibinfo{booktitle}{\emph{Proc. of IMIS}}.
\newblock


\bibitem[Schulzrinne et~al\mbox{.}(2003)]%
        {rfc3550}
\bibfield{author}{\bibinfo{person}{Henning Schulzrinne},
  \bibinfo{person}{Steven Casner}, \bibinfo{person}{R Frederick}, {and}
  \bibinfo{person}{Van Jacobson}.} \bibinfo{year}{2003}\natexlab{}.
\newblock \bibinfo{title}{RFC3550: RTP: A transport protocol for real-time
  applications}.
\newblock
\newblock


\bibitem[Sharma et~al\mbox{.}(2022)]%
        {sharma2022benchmarks}
\bibfield{author}{\bibinfo{person}{Ranya Sharma}, \bibinfo{person}{Tarun
  Mangla}, \bibinfo{person}{James Saxon}, \bibinfo{person}{Marc Richardson},
  \bibinfo{person}{Nick Feamster}, {and} \bibinfo{person}{Nicole~P Marwell}.}
  \bibinfo{year}{2022}\natexlab{}.
\newblock \showarticletitle{Benchmarks or Equity? A New Approach to Measuring
  Internet Performance}.
\newblock \bibinfo{journal}{\emph{A New Approach to Measuring Internet
  Performance (August 3, 2022)}} (\bibinfo{year}{2022}).
\newblock


\bibitem[Sharma et~al\mbox{.}(2023)]%
        {githubrepo}
\bibfield{author}{\bibinfo{person}{Taveesh Sharma}, \bibinfo{person}{Tarun
  Mangla}, \bibinfo{person}{Arpit Gupta}, \bibinfo{person}{Junchen Jiang},
  {and} \bibinfo{person}{Nick Feamster}.} \bibinfo{year}{2023}\natexlab{}.
\newblock \bibinfo{booktitle}{\emph{vcaml}}.
\newblock
\urldef\tempurl%
\url{https://github.com/noise-lab/vcaml}
\showURL{%
\tempurl}
\newblock
\shownote{Estimating WebRTC Video QoE Metrics Without Using Application
  Headers}.


\bibitem[Sonchack et~al\mbox{.}(2018)]%
        {star-flow}
\bibfield{author}{\bibinfo{person}{John Sonchack}, \bibinfo{person}{Oliver
  Michel}, \bibinfo{person}{Adam~J Aviv}, \bibinfo{person}{Eric Keller}, {and}
  \bibinfo{person}{Jonathan~M Smith}.} \bibinfo{year}{2018}\natexlab{}.
\newblock \showarticletitle{Scaling Hardware Accelerated Network Monitoring to
  Concurrent and Dynamic Queries With $\{$* Flow$\}$}. In
  \bibinfo{booktitle}{\emph{2018 USENIX Annual Technical Conference (USENIX ATC
  18)}}. \bibinfo{pages}{823--835}.
\newblock


\bibitem[Uberti et~al\mbox{.}(2023)]%
        {uberti2023rfc}
\bibfield{author}{\bibinfo{person}{J Uberti}, \bibinfo{person}{C Jennings},
  {and} \bibinfo{person}{S Murillo}.} \bibinfo{year}{2023}\natexlab{}.
\newblock \bibinfo{title}{RFC 9335: Completely Encrypting RTP Header Extensions
  and Contributing Sources}.
\newblock
\newblock


\bibitem[Varvello et~al\mbox{.}(2022)]%
        {varvello2022performance}
\bibfield{author}{\bibinfo{person}{Matteo Varvello}, \bibinfo{person}{Hyunseok
  Chang}, {and} \bibinfo{person}{Yasir Zaki}.} \bibinfo{year}{2022}\natexlab{}.
\newblock \showarticletitle{Performance characterization of videoconferencing
  in the wild}. In \bibinfo{booktitle}{\emph{Proceedings of the 22nd ACM
  Internet Measurement Conference}}. \bibinfo{pages}{261--273}.
\newblock


\bibitem[Yan et~al\mbox{.}(2015)]%
        {Yan2015EnablingQL}
\bibfield{author}{\bibinfo{person}{Suying Yan}, \bibinfo{person}{Yuchun Guo},
  \bibinfo{person}{Yishuai Chen}, \bibinfo{person}{Feng Xie},
  \bibinfo{person}{Chenguang Yu}, {and} \bibinfo{person}{Y. Liu}.}
  \bibinfo{year}{2015}\natexlab{}.
\newblock \showarticletitle{Enabling QoE Learning and Prediction of WebRTC
  Video Communication in WiFi Networks}.
\newblock
\urldef\tempurl%
\url{https://api.semanticscholar.org/CorpusID:30220021}
\showURL{%
\tempurl}


\bibitem[Yan et~al\mbox{.}(2017)]%
        {yan2017enabling}
\bibfield{author}{\bibinfo{person}{Suying Yan}, \bibinfo{person}{Yuchun Guo},
  \bibinfo{person}{Yishuai Chen}, \bibinfo{person}{Feng Xie},
  \bibinfo{person}{Chenguang Yu}, {and} \bibinfo{person}{Yong Liu}.}
  \bibinfo{year}{2017}\natexlab{}.
\newblock \showarticletitle{Enabling QoE learning and prediction of WebRTC
  video communication in WiFi networks}. In
  \bibinfo{booktitle}{\emph{Proceedings of the ICC}},
  Vol.~\bibinfo{volume}{2017}.
\newblock


\end{thebibliography}
\appendix
\thispagestyle{empty}

\setcounter{figure}{0}
\renewcommand\thefigure{A.\arabic{figure}}

\setcounter{table}{0}
\renewcommand\thetable{A.\arabic{table}}

\section{Statement of Ethics}
The real-world network traces used in this paper are collected after
obtaining approvals from our Institutional Review Board (IRB).
We prioritize the protection of user privacy and take extensive measures
to ensure it. Our deployment setup solely permits the
collection of active
measurement data from participants' homes; we can not monitor any user
network traffic. \rev{}{More specifically, the Raspberry Pi (RPi) devices used for this study are connected to the home router using a wired connection like any other device. We do not sit in the middle of the user device and the home router.} Additionally, we remove any personally
identifiable information, such as physical address and demographics,
before analyzing the collected data. 

\rev{}{The network trace data that we make public corresponds to the VCA calls between the Raspberry Pi and the cloud endpoint. As an additional privacy measure, the IP addresses of both these endpoints have been hashed in the network traces as well as the JSON files obtained via {\tt webrtc-internals}}. The remaining datasets used
in this paper are collected within controlled lab setting and do not
pose any privacy-related issues. 

\section{Methodology}
\begin{algorithm}[H]
  \caption{An algorithm for VCA frame boundary estimation using IP/UDP headers
    only}\label{algo:ip-heuristic}
  \begin{algorithmic}
    \Require $packets,~\Delta_{size}^{max},~N^{max}$
    \Ensure $frames$
    \State $f \gets 0$
    \State $frames \gets \{\}$
    \For {$p$ in $packets$}
    \State $assigned \gets False$
    \For {$p'$ in previously seen $N^{max}$ $packets$}
    \If{$|p'.size$ - $p.size| \leq \Delta_{size}^{max}$}
    \State $frames~[p] \gets frames~[p']$
    \State $assigned \gets True$
    \State \textbf{break}
    \EndIf
    \EndFor
    \If{$assigned = True$}
    \State $f \gets f + 1$
    \State $frames~[p] \gets f$
    \EndIf
    \EndFor
  \end{algorithmic}
\end{algorithm}

\section{Datasets}

\subsection{Data Description}
\begin{figure}[h]
	\centering
			\begin{subfigure}[t!]{0.65\linewidth}
		\includegraphics[width=\linewidth,keepaspectratio]{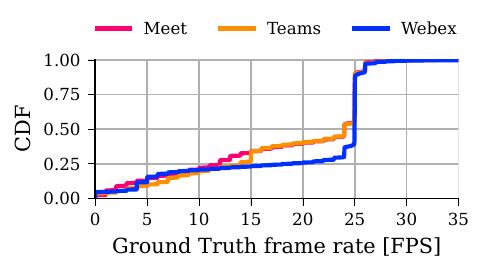}
		\caption{Frames per second}
		\label{subfig:fps_gt}
			\end{subfigure}
\begin{subfigure}[t!]{0.65\linewidth}
		\centering
		\includegraphics[width=\linewidth]{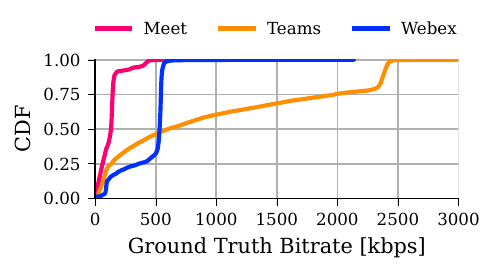}
		\caption{Video bitrate}
		\label{subfig:bitrate_gt}
	\end{subfigure}
	\begin{subfigure}[t!]{0.65\linewidth}
		\centering
		\includegraphics[width=\linewidth]{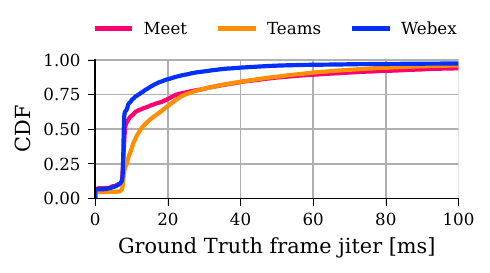}
		\caption{Frame Jitter}
		\label{subfig:std_if_gt}
	\end{subfigure}
\caption{CDF of ground truth QoE metrics for in-lab data}
\label{fig:qoe-cdf}
\end{figure}

\begin{figure}[t!]
	\centering
			\begin{subfigure}[t!]{0.65\linewidth}
		\includegraphics[width=\linewidth,keepaspectratio]{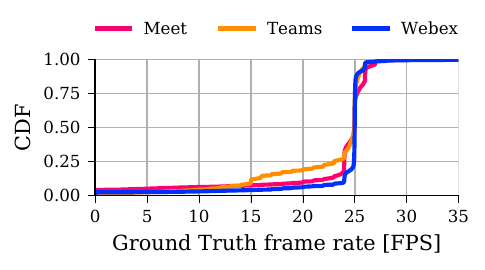}
		\caption{Frames per second}
		\label{subfig:fps_gt_rw}
			\end{subfigure}
\begin{subfigure}[t!]{0.65\linewidth}
		\centering
		\includegraphics[width=\linewidth]{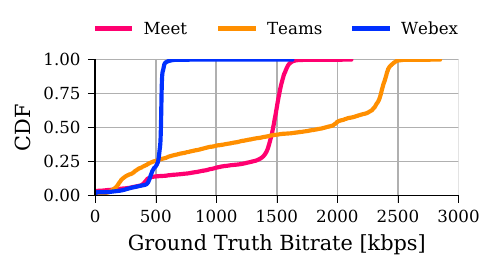}
		\caption{Video bitrate}
		\label{subfig:bitrate_gt_rw}
	\end{subfigure}
	\begin{subfigure}[t!]{0.65\linewidth}
		\centering
		\includegraphics[width=\linewidth]{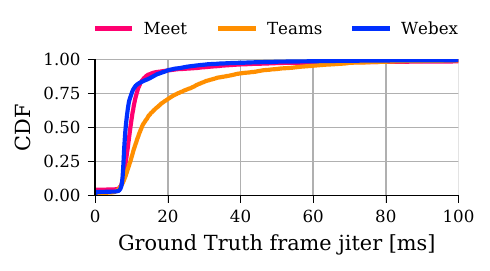}
		\caption{Frame Jitter}
		\label{subfig:std_if_gt_rw}
	\end{subfigure}
\caption{CDF of ground truth QoE metrics for real-world data}
\label{fig:qoe-cdf-rw}
\end{figure}

Figure~\ref{fig:qoe-cdf} and Figure~\ref{fig:qoe-cdf-rw} show the CDF of ground
truth QoE metrics for in-lab and real-world datasets respectively.

\begin{table}[H]
  \centering
  \small
  \begin{tabular}{|c|cc|c|}
  \hline
  \multirow{2}{*}{\textbf{Actual}} & \multicolumn{2}{c|}{\textbf{Prediction}} & \multirow{2}{*}{\textbf{Total}} \\ \cline{2-3}
                                   & \multicolumn{1}{c|}{Non-Video}  & Video  &                                 \\ \hline
  Non-video                        & \multicolumn{1}{c|}{98.2\%}     & 1.8\%  & 50,799                          \\ \hline
  Video                            & \multicolumn{1}{c|}{0\%}        & 100\%  & 946,769                         \\ \hline
  \end{tabular}
  \caption{Webex Media classification accuracy for in-lab data}
  \label{tab:webex-media}
  \end{table}

\begin{table}[H]
\centering
\small
\begin{tabular}{|c|cc|c|}
\hline
\multirow{2}{*}{\textbf{Actual}} & \multicolumn{2}{c|}{\textbf{Prediction}} & \multirow{2}{*}{\textbf{Total}} \\ \cline{2-3}
                                 & \multicolumn{1}{c|}{Non-Video}  & Video  &                                 \\ \hline
Non-video                        & \multicolumn{1}{c|}{98.5\%}     & 1.5\%    & 378,249                         \\ \hline
Video                            & \multicolumn{1}{c|}{0\%}        & 100\%  & 1,818,689                       \\ \hline
\end{tabular}
\caption{Teams Media classification accuracy for in-lab data}
\label{tab:teams-media}
\end{table}

\section{Evaluation}

\subsection{In-lab Data}

\subsubsection{Media classification accuracy}
Table~\ref{tab:teams-media} and~\ref{tab:webex-media} show the media
classification accuracy of \teams and \webex, respectively, using only
IP/UDP headers.

\subsubsection{Frame rate}

\begin{figure}[H] \begin{center}
  \includegraphics[width=0.65\linewidth]{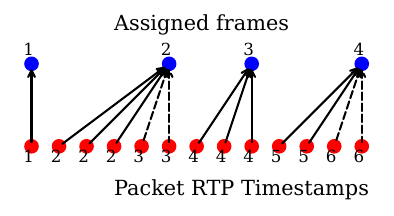}
\end{center}
\caption{A plot showing frame assignments by the \ipnonml
  approach over a 1-second window for Meet. The solid arrows represent correct
  frame assignments while the dotted arrows represent incorrect ones.}
\label{fig:teams-asgn}
\end{figure}
Figure~\ref{fig:teams-asgn} illustrates a case of frame coalescing from one of the \teams sessions.
The red dots represent sequence of packets over time with their
respective RTP timestamp, while the blue dots show the frame
assignment by the \ipnonml. Packets with RTP timestamp 2 and 3 have
a size of 1022 bytes and 1020 bytes, respectively, leading to these
packets grouped into a single frame. Similar is the case for packets
with RTP timestamp 5 and 6.

\begin{figure*}[t!]
  \begin{subfigure}{0.3\linewidth} \centering
    \includegraphics[width=\linewidth]{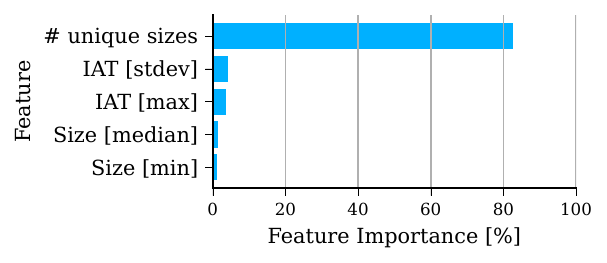}
    \caption{\meet} \label{subfig:fimp-ip-meet}
  \end{subfigure}%
  \begin{subfigure}{0.3\linewidth} \centering
    \includegraphics[width=\linewidth]{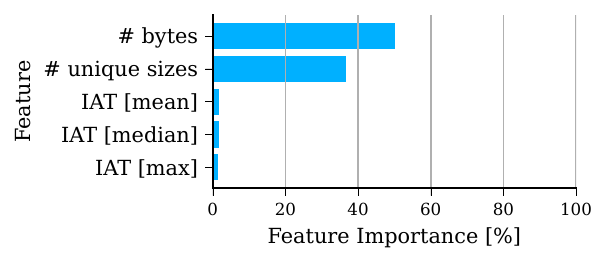}
    \caption{\webex} \label{subfig:fimp-ip-webex}
  \end{subfigure}
  \caption{Top-5 features along with importance scores for frame rate
    estimation across the
    three VCAs for the \ipml method} \label{fig:fps-fimp}
\end{figure*}

\begin{figure*}[t!]
  \begin{subfigure}{0.3\linewidth} \centering
    \includegraphics[width=\linewidth]{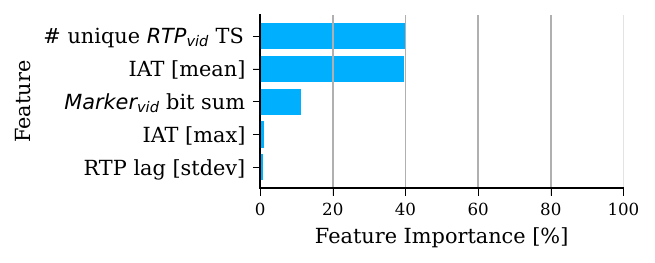}
    \caption{\meet} \label{subfig:fimp-ip-meet}
  \end{subfigure}%
  \begin{subfigure}{0.3\linewidth} \centering
    \includegraphics[width=\linewidth]{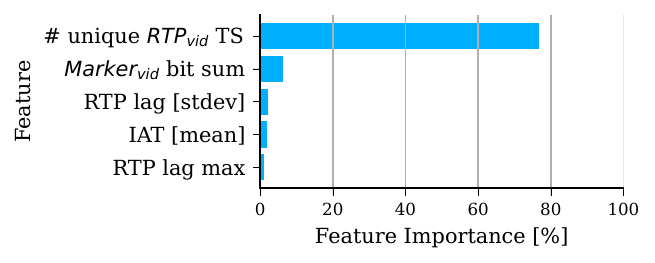}
    \caption{\teams} \label{subfig:fimp-ip-webex}
  \end{subfigure}
  \begin{subfigure}{0.3\linewidth} \centering
    \includegraphics[width=\linewidth]{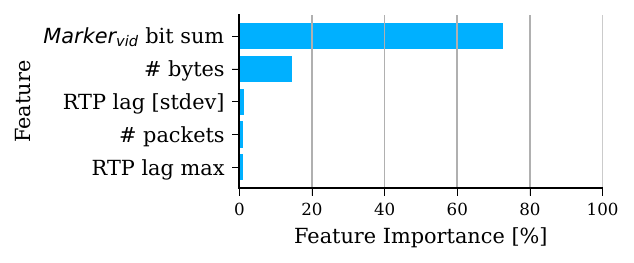}
    \caption{\webex} \label{subfig:fimp-ip-webex}
  \end{subfigure}
  \caption{Top-5 features along with importance scores for frame rate
    estimation across the
    three VCAs for the \rtpml method} \label{fig:fps-fimp-rtp}
\end{figure*}

\paragraph{Feature Importance}. Figure~\ref{fig:fps-fimp}
and~\ref{fig:fps-fimp-rtp} show the
feature importance plots for IP/UDP ML and RTP ML methods,
respectively.

\begin{figure*}[t!]
  \begin{subfigure}{0.3\linewidth}
    \includegraphics[width=1.0\linewidth]{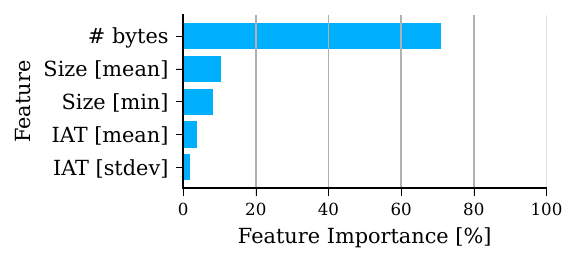}
    \caption{\meet}
    \label{subfig:bitrate-fimp-ip}
  \end{subfigure}%
  \begin{subfigure}{0.3\linewidth}
    \includegraphics[width=1.0\linewidth]
    {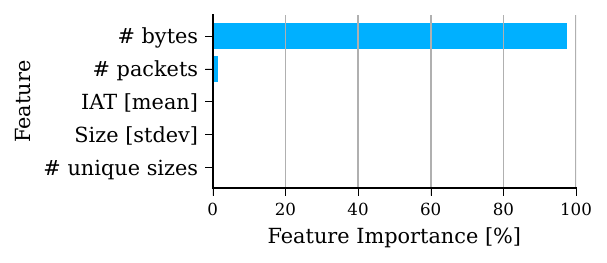}
    \caption{\teams}
    \label{subfig:bitrate-fimp-rtp}
  \end{subfigure}%
  \caption{Top-5 features along with feature importance scores for bitrate
    estimation using the
    \ipml method.} \label{fig:bitrate-fimp-ip}
\end{figure*}

\begin{figure*}[t!]
  \begin{subfigure}{0.3\linewidth}
    \includegraphics[width=1.0\linewidth]{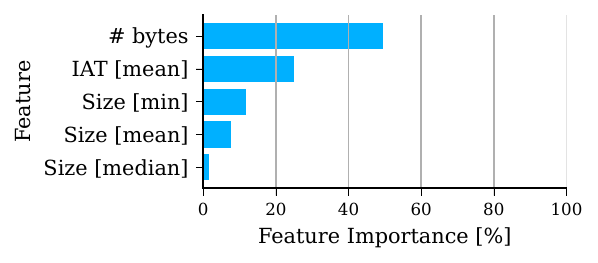}
    \caption{\meet}
    \label{subfig:bitrate-fimp-ip}
  \end{subfigure}%
  \begin{subfigure}{0.3\linewidth}
    \includegraphics[width=1.0\linewidth]
    {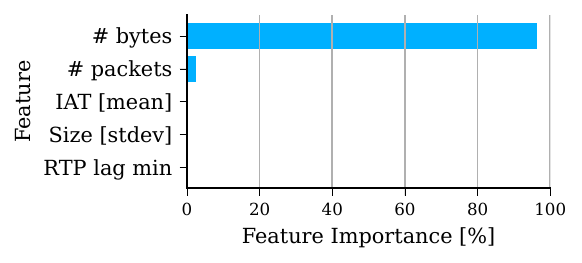}
    \caption{\teams}
    \label{subfig:bitrate-fimp-rtp}
  \end{subfigure}%
  \begin{subfigure}{0.3\linewidth}
    \includegraphics[width=1.0\linewidth]
    {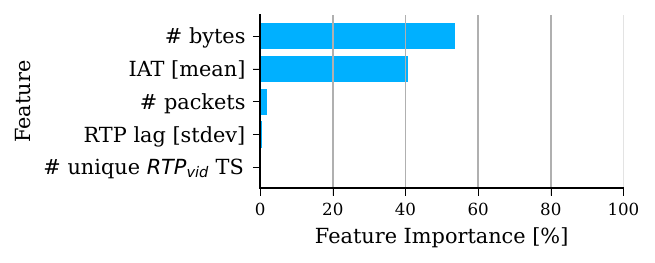}
    \caption{\webex}
    \label{subfig:bitrate-fimp-rtp}
  \end{subfigure}%
  \caption{Top-5 features along with feature importance scores for bitrate
    estimation using the
    \rtpml method.} \label{fig:bitrate-fimp-rtp}
\end{figure*}

\subsubsection{Video bitrate}
\paragraph{Feature Importance}. Figure~\ref{fig:bitrate-fimp-ip}
and~\ref{fig:bitrate-fimp-rtp} show the
feature importance plots for IP/UDP ML and RTP ML methods,
respectively.

\begin{figure*}[t!]
  \begin{subfigure}{0.3\linewidth}
    \includegraphics[width=\linewidth]
    {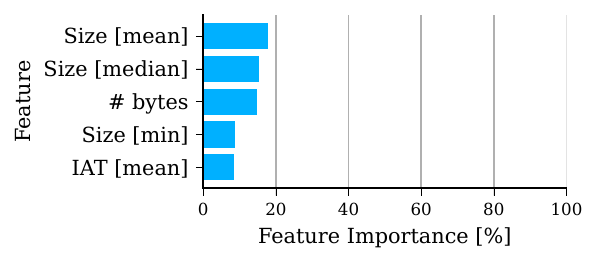}
    \caption{\meet}
    \label{subfig:resolution-fimp-ip}
  \end{subfigure}%
  \begin{subfigure}{0.3\linewidth}
    \includegraphics[width=\linewidth]
    {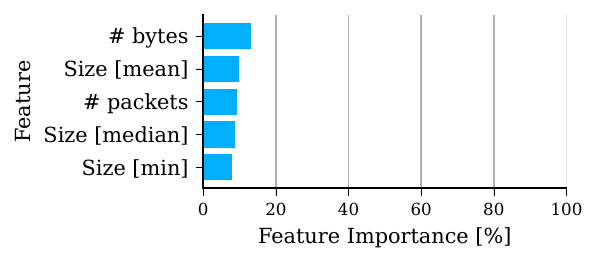}
    \caption{\teams}
    \label{subfig:resolution-fimp-ip}
  \end{subfigure}%
  \caption{Top-5 features along with feature importance scores for
    resolution
    estimation using the
    \ipml method.} \label{fig:resolution-fimp}
\end{figure*}

\begin{figure*}[t!]
  \begin{subfigure}{0.3\linewidth}
    \includegraphics[width=\linewidth]
    {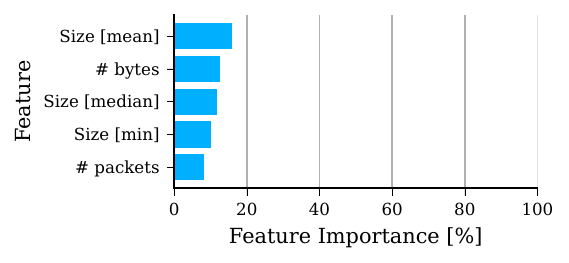}
    \caption{\meet}
    \label{subfig:resolution-fimp-rtp}
  \end{subfigure}%
  \begin{subfigure}{0.3\linewidth}
    \includegraphics[width=\linewidth]
    {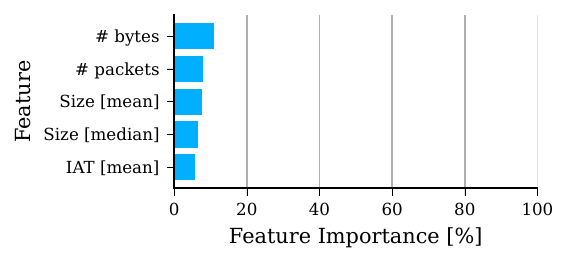}
    \caption{\teams}
    \label{subfig:resolution-fimp-rtp}
  \end{subfigure}%
  \begin{subfigure}{0.3\linewidth}
    \includegraphics[width=\linewidth]
    {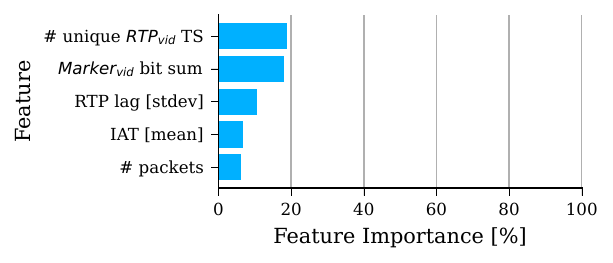}
    \caption{\webex}
    \label{subfig:resolution-fimp-rtp}
  \end{subfigure}
  \caption{Top-5 features along with feature importance scores for
    resolution
    estimation using the
    \rtpml method.} \label{fig:resolution-fimp-rtp}
\end{figure*}

\subsubsection{Frame Resolution}
\paragraph{Feature Importance}. Figure~\ref{fig:resolution-fimp}
and~\ref{fig:resolution-fimp-rtp} show the
feature importance plots for IP/UDP ML and RTP ML methods,
respectively.

\subsection{Real-world Data}
\begin{table}[t!]
  \small
  \centering
  \begin{tabular}{|c|ccc|c|}
    \hline
    \multirow{2}{*}{\textbf{Actual}} & \multicolumn{3}{c|}{\textbf{Predicted}} & \multirow{2}{*}{\textbf{Total}}                  \\ \cline{2-4}
                                     & \multicolumn{1}{c|}{Low}                & \multicolumn{1}{c|}{Medium}     & High    &      \\ \hline
    Low                              & \multicolumn{1}{c|}{90.23\%}            & \multicolumn{1}{c|}{5.58\%}     & 4.19\%  & 573  \\ \hline
    Medium                           & \multicolumn{1}{c|}{14.32\%}            & \multicolumn{1}{c|}{30.87\%}    & 54.81\% & 447  \\ \hline
    High                             & \multicolumn{1}{c|}{0.89\%}             & \multicolumn{1}{c|}{3.34\%}     & 95.77\% & 2576 \\ \hline
  \end{tabular}
  \caption{The normalized confusion matrix for resolution predictions
    by \ipml model for Teams on real-world data. The percentages indicate the accuracy of our
    predictions for each frame height.}
  \label{tab:resolution-cm-rw}
\end{table}

\subsubsection{Resolution}
Table~\ref{tab:resolution-cm-rw} shows the \ipml confusion matrix for
resolution prediction for \teams
on real-world data.

\subsection{Model Transferability}

\begin{table}[t!]
  \small
  \centering
  \begin{tabular}{|c|ccc|}
    \hline
    \multirow{2}{*}{\textbf{Method}} & \multicolumn{3}{c|}{\textbf{VCA}}                                                         \\ \cline{2-4}
                                     & \multicolumn{1}{c|}{\textbf{Meet}} & \multicolumn{1}{c|}{\textbf{Teams}} & \textbf{Webex} \\ \hline
    \ipml                            & \multicolumn{1}{c|}{889.93}        & \multicolumn{1}{c|}{114.06}         & 29.53          \\ \hline
    \rtpml                           & \multicolumn{1}{c|}{793.86}        & \multicolumn{1}{c|}{167.18}         & 29.22          \\ \hline
  \end{tabular}
  \caption{Bitrate MAE results after using lab-trained models to predict real-world MAE}
  \label{tab:bitrate-transfer}
\end{table}

\begin{table}[t!]
  \small
  \centering
  \begin{tabular}{|c|ccc|}
    \hline
    \multirow{2}{*}{\textbf{Method}} & \multicolumn{3}{c|}{\textbf{VCA}}                                                         \\ \cline{2-4}
                                     & \multicolumn{1}{c|}{\textbf{Meet}} & \multicolumn{1}{c|}{\textbf{Teams}} & \textbf{Webex} \\ \hline
    \ipml                            & \multicolumn{1}{c|}{89.74}         & \multicolumn{1}{c|}{64.36}          & 29.78          \\ \hline
    \rtpml                           & \multicolumn{1}{c|}{30.31}         & \multicolumn{1}{c|}{19.87}          & 95.43          \\ \hline
  \end{tabular}
  \caption{Frame Jitter MAE results after using lab-trained models to predict real-world MAE}
  \label{tab:jitter-transfer}
\end{table}

Table~\ref{tab:bitrate-transfer} and~\ref{tab:jitter-transfer} show the
MAE of models trained using in-lab data and tested on real-world data
for video bitrate and frame jitter, respectively.

\subsection{Effect of Network Conditions}

\begin{table*}[t!]
  \small
  \centering
  \begin{tabular}{|c|c|c|c|}
    \hline
    \textbf{Impairment} & \textbf{Throughput [kbps]}                    &
    \textbf{Delay [ms]} & \textbf{Packet Loss}                            \\ \hline
    Mean Throughput     & $\mu$ : [100, 200, 500,1000, 2000, 4000],
    $\sigma$: 0         & $\mu$: 50, $\sigma$: 0                        &
    0\%                                                                   \\ \hline
    Throughput stdev.   & $\mu$: 1500, $\sigma$: [0, 100, 200, 500,
    1000, 1500]         & $\mu$: 50, $\sigma$: 0                        &
    0\%                                                                   \\ \hline
    Mean Latency        & $\mu$: 1500, $\sigma$: 0
                        & $\mu$: [50, 100, 200, 300, 400, 500],
    $\sigma$: 0         & 0\%                                             \\ \hline
    Latency stdev.      & $\mu$: 1500, $\sigma$: 0
                        & $\mu$: 50, $\sigma$: [10, 20, 30, 40, 50, 60,
    70, 80, 90, 100]    & 0\%                                             \\ \hline
    Packet Loss \%      & $\mu$: 1500, $\sigma$: 0
                        & $\mu$: 50, $\sigma$: 0
                        & [1, 2, 5, 10, 15, 20]\%                         \\ \hline
  \end{tabular}
  \caption{Different impairment profiles used for network sensitivity tests.
    Square brackets indicate a variation across different calls. $\mu$ and $\sigma$
    denote mean and standard deviation respectively.}
  \label{table:net-conditions}
\end{table*}

Table~\ref{table:net-conditions} summarizes the synthetic network
conditions emulated to study the effect of network conditions on the
accuracy of ML models.

\subsection{Effect of \ipnonml packet lookback}

\begin{figure}[H]
  \begin{center}
    \includegraphics[width=0.75\linewidth]{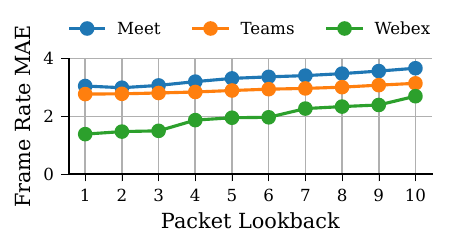}
  \end{center}
  \caption{Variation of frame rate MAE with \ipnonml packet lookback parameter}
  \label{fig:lookback_tuning}
\end{figure}

The \ipnonml packet lookback parameter was tuned on a sample of 50 in-lab
traces each for Meet, Teams and Webex. Figure~\ref{fig:lookback_tuning} shows
the variation of frame rate MAE with the number of packets we look back to match
a packet with already assembled frames. For Webex we see a clear increasing
trend, while for Meet and Teams we observe \rev{minimas}{minima} at lookbacks of 3 and 2
respectively. Webex has an optimal lookback of 1 because 99.70\% frames have a
maximum intra-frame size difference of 2 bytes, and 99.38\% of the frames are of
size less than or equal to 3 packets. Our algorithm is thus able to merge
similarly sized frames together by not looking too far back. For Teams, even
though 98.56\% of the frames have an intra-frame size difference of 2 bytes,
only 43.82\% have a size less than or equal to 3 packets. Thus, a greater
lookback is required to merge similarly sized packets together. For Meet, these
percentages are slightly lower than Webex (95.73\% and 95.18\%), thus the
optimal lookback is 2 packets.

\end{sloppypar}
\end{document}